\documentclass[]{pasj01}
\usepackage{color}
\usepackage{bm}
\usepackage{ulem}
\usepackage{arydshln}
\usepackage{lineno}
\usepackage{multirow}

\begin{document} 
\Received{yyyy/mm/dd}
\Accepted{yyyy/mm/dd}

\newcommand*{\ta}[1]{\textcolor{cyan}{#1}}
\newcommand*{\kk}[1]{\textcolor{magenta}{#1}}
\newcommand*{\comment}[1]{\textcolor{blue}{#1}}
\newcommand*{\rev}[1]{\textcolor{black}{#1}}

\newcommand{\norv}[1]{{\textcolor{blue}{{\bf Nobu:}#1}}}
\newcommand{\HA}[1]{{\textcolor{red}{{\bf HA:}#1}}}

\newcommand{\CenterRow}[2]{
  \dimen0=\ht\strutbox%
  \advance\dimen0\dp\strutbox%
  \multiply\dimen0 by#1%
  \divide\dimen0 by2%
  \advance\dimen0 by-.5\normalbaselineskip%
  \raisebox{-\dimen0}[0pt][0pt]{#2}}

\title{{Diffuse radio source candidate in CIZA J1358.9-4750}}

\author{Kohei \textsc{KURAHARA}\altaffilmark{1,*}%
}
\altaffiltext{1}{Mizusawa VLBI Observatory, National Astronomical Observatory of Japan, 2-21-1 Osawa, Mitaka, Tokyo 181-8588, Japan}
\email{kohei.kurahara@nao.ac.jp}

\author{Takuya \textsc{AKAHORI}\altaffilmark{1,2}}
\altaffiltext{2}{Operation Division, Square Kilometre Array Observatory, Lower Withington, Macclesfield, Cheshire SK11 9FT, UK}

\author{Ruta \textsc{KALE}\altaffilmark{3}}
\altaffiltext{3}{National Centre for Radio Astrophysics, Tata Institute of Fundamental Research, S. P. Pune University Campus, Ganeshkhind, Pune 411007, India}

\author{Hiroki \textsc{AKAMATSU}\altaffilmark{4}}
\altaffiltext{4}{SRON Netherlands Institute for Space Research, Niels Bohrweg 4, 2333 CA Leiden, The Netherlands}

\author{Yutaka \textsc{FUJITA}\altaffilmark{5}}
\altaffiltext{5}{Department of Physics, Graduate School of Science, Tokyo Metropolitan University, 1-1 Minami-Osawa, Hachioji-shi, Tokyo 192-0397, Japan}

\author{Liyi \textsc{GU}\altaffilmark{4}}

\author{Huib \textsc{INTEMA}\altaffilmark{6}}
\altaffiltext{6}{Leiden Observatory, Leiden University, Niels Bohrweg 2, 2333 CA, Leiden, The Netherlands}

\author{Kazuhiro \textsc{NAKAZAWA}\altaffilmark{7}}
\altaffiltext{7}{The Kobayashi-Maskawa Institute for the Origin of Particles and the Universe (or KMI), Nagoya University, Furo-cho, Chikusa-ku, Nagoya, 464-8602, Japan}

\author{Nobuhiro \textsc{OKABE}\altaffilmark{8,9}}
\altaffiltext{8}{Department of Physical Science, Hiroshima University, 1-3-1 Kagamiyama, Higashi-Hiroshima,
Hiroshima 739-8526, Japan}
\altaffiltext{9}{Hiroshima Astrophysical Science Center, Hiroshima University, 1-3-1 Kagamiyama, Higashi-Hiroshima,
Hiroshima 739-8526, Japan}

\author{Yuki \textsc{OMIYA}\altaffilmark{10}}
\altaffiltext{10}{Departure of Physics, Nagoya University, Furo-cho, Chikusa-ku, Nagoya, Aichi 464-8601, Japan}

\author{Viral  \textsc{PAREKH}\altaffilmark{11,12}}
\altaffiltext{11}{Centre for Radio Astronomy Techniques and Technologies, Department of Physics and Electronics, Rhodes University, PO Box 94, Makhanda, 6140, South Africa}
\altaffiltext{12}{South African Radio Astronomy Observatory (SARAO), 2 Fir Street, Black River Park, Observatory, Cape Town 7925, South Africa}

\author{Timothy \textsc{SHIMWELL}\altaffilmark{13,14}}
\altaffiltext{13}{Leiden Observatory, Leiden University, PO Box 9513, NL-2300 RA Leiden, The Netherlands}
\altaffiltext{14}{ASTRON, the Netherlands Institute for Radio Astronomy, Oude Hoogeveensedjjk 4, 7991 PD Dwingeloo, The Netherlands}

\author{Motokazu \textsc{TAKIZAWA}\altaffilmark{15}}
\altaffiltext{15}{Department of Physics, Yamagata University, Kojirakawa-machi 1-4-12, Yamagata, Yamagata 990-8560, Japan}

\author{Reinout \textsc{Van WEEREN}\altaffilmark{16}}
\altaffiltext{16}{Leiden Observatory, Leiden University, Niels Bohrweg 2, 2300 RA Leiden, The Netherlands}



\maketitle

\begin{abstract}
We report on results of our upgraded Giant Metrewave Radio Telescope (uGMRT) observations for an early-stage merging galaxy cluster, CIZA J1358.9-4750 (CIZA1359), in Band-3 (300--500 MHz). We achieved the image dynamic range of $\sim 38,000$ using the direction dependent calibration and found a candidate of diffuse radio emission at 4~$\sigma_{rms}$ significance. The flux density of the candidate at 400~MHz, $24.04 \pm 2.48$~mJy, is significantly positive compared to noise, where its radio power, $2.40 \times 10^{24}$~W~Hz$^{-1}$, is consistent with those of typical diffuse radio sources of galaxy clusters. The candidate is associated with a part of the X-ray shock front at which the Mach number reaches its maximum value of $\mathcal{M}\sim 1.7$. The spectral index ($F_\nu \propto \nu^{\alpha}$) of the candidate, $\alpha = - 1.22 \pm 0.33$,  is in agreement with an expected value derived from the standard diffusive shock acceleration (DSA) model. But such a low Mach number with a short acceleration time would require seed cosmic-rays supplied from active galactic nucleus (AGN) activities of member galaxies, as suggested in some other clusters. Indeed, we found seven AGN candidates inside the diffuse source candidate. Assuming the energy equipartition between magnetic fields and cosmic-rays, the magnetic field strength of the candidate was estimated to be $2.1~\mu$G. We also find head-tail galaxies and radio phoenixes or fossils near the CIZA1359.
\end{abstract}
\section{Introduction}

The largest self-gravitating systems in the universe, galaxy clusters, are several Mpc in size and $10^{14-15}$ M$_{\odot}$ in mass and are known to contain hot ($10^{7-8}$ K) intracluster medium (ICM). This thermal energy is thought to be converted from huge gravitational energy of the large-scale structure through the bottom-up structure formation. Pairs of two sub-clusters in close location are thought to be colliding each other (e.g., \cite{2007PhR...443....1M} for a review), called merging clusters, and they are the sites of this energy conversion. A major energy-conversion mechanism is believed to be a shock wave formed during the merger. However, detailed physical mechanisms of shock wave such as particle acceleration, magnetic-field amplification, and turbulence generation are longstanding questions in astrophysics.

Shock wave in the ICM is often identified from X-ray observations of ICM's density and/or temperature jumps. The shock is also found from radio observations of synchrotron radiation, called radio relics, emitted from the shock-accelerated cosmic-ray electrons (e.g., \cite{2012A&ARv..20...54F, 2018PASJ...70R...2A, 2019SSRv..215...16V} for reviews). The Fermi 1st order acceleration, namely the diffusive shock acceleration (DSA, \cite{1987PhR...154....1B}), is thought to be one of the plausible theories for the particle acceleration. Meanwhile, there are other classes of diffuse radio emission in galaxy clusters, radio halos, mini-halos, and radio bridges. Some of them are thought to be formed by turbulence based on the Fermi 2nd order acceleration \citep{2001MNRAS.320..365B, 2003ApJ...584..190F}. Radio bridge is a relatively new class of diffuse radio emission, which was found at the linked region of early-stage merging clusters \citep{2019Sci...364..981G, 2020PhRvL.124e1101B, 2020MNRAS.499L..11B}. Since turbulence acceleration suffers from the acceleration efficiency, seed cosmic-rays supplied from AGN jets of member galaxies were proposed for the radio bridges \citep{2020PhRvL.124e1101B}.

Another important factor for thermal evolution of the ICM is AGN jets launched from the supermassive black holes of member galaxies. In the last decades, AGN jet is thought to be a promising source to solve the so-called cooling flow problem (\cite{1994ARA&A..32..277F} for a review), while the co-existence of cooling gas and AGN jets in Phoenix galaxy cluster \citep{2020PASJ...72...33K, 2020PASJ...72...62A} raise a new question on AGN feedback. A bent AGN jet in Abell 3376 indicated a tight connection between the jet and the coherent magnetic field at the cold front of the cluster \citep{2021Natur.593...47C}. The spectral index distribution exhibits a plateau near the bending point, suggesting re-acceleration of cosmic rays likely by magnetic reconnection. Recently, AGN jet is more recognized as a source of cosmic rays in the ICM. AGN jets connecting to radio relics are found in, for example, Abell~3411 \citep{2017NatAs...1E...5V} and Abell 3376 \citep{2022PASJ..tmp...40C}. In Abell~3411, the spectral index changes continuously along the radio structure, indicating spectral aging caused by cosmic-ray electron cooling. Therefore, a detailed study of radio sources in galaxy clusters can provide a new knowledge, such as particle acceleration, in addition to understanding the evolution of the sources themselves.

CIZA J1358.9-4750 (CIZA1359) is thought to be one of the only several known early-stage merging galaxy clusters. This object is a Clusters in the Zone of Avoidance (CIZA) survey target, thus CIZA1359 is found relatively close to the Galactic Plane \citep{2007ApJ...662..224K}. These basic information are summarized in table \ref{tab:1}. \citet{2015PASJ...67...71K} performed a detailed analysis of the Suzaku X-ray observation and found a discontinuous high-temperature region in the linked region between the two X-ray peaks of subclusters. They suggested that the high-temperature region was formed by the merger shock wave passing along the merger axis. \citet{2022arXiv221002145O} further studied X-ray properties of CIZA1359 and found another shock front at the northern edge of the hot region. These studies also suggest that the merger axis is off the plane of the sky.

CIZA1359 is relatively nearby at redshift $z=0.074$, which is convenient for studying its detail, and is expected to be investigated in more detail. In this paper, we report on the results of the upgraded Giant Metrewave Radio Telescope (uGMRT) observation of CIZA1359, with the aim of detecting any diffuse emission of CIZA1359. In Section 2 we describes the details of the uGMRT observations and the data reduction, and in Section 3 we present some obtained radio image and spectral index map. In Section 4, we discuss on relic candidate of CIZA1359. We have used cosmological parameters $H_0 = 70~{\rm km~s^{-1}~Mpc^{-1}}$, $\Omega_M = 0.3$ and $\Omega_\Lambda = 0.7$ in this work.

\section{Observation and Data reduction}
\subsection{The Observation}

\begin{table}[tp]
\tbl{Basic parameters of CIZA1359}{%
\begin{tabular}{llc} \hline
Parameter & Value & reference \\ \hline
RA (J2000)& $13^h58^m40^s$ & [1]\\ 
Dec (J2000)& $-47^d46^m00^s$ & [1] \\
Redshift & 0.0740 & [1] \\
$f_X[0.1\sim 2.4{\rm keV}]$ & $20.89 \times 10^{-12}~{\rm erg~cm^{-1}}$  & [1] \\
$L_X[0.1\sim 2.4{\rm keV}]$ & $4.88 \times 10^{44}~{\rm erg~s^{-1}}$ & [1] \\
kT (keV) &\begin{tabular}{l} $5.6 \pm 0.2 {\rm keV}$ (south-east)\\ $4.6 \pm 0.2 {\rm keV}$ (north-west) \end{tabular} & [2] \\ \hline
\end{tabular}}\label{tab:1}
\begin{tabnote}
[1] \cite{2007ApJ...662..224K}; [2] \cite{2015PASJ...67...71K}
\end{tabnote}
\end{table}

We conducted uGMRT Band~3 (300 -- 500~MHz) observations of CIZA1359 (the project code 39\_045). Both narrow- and wide-band modes were adopted. The center frequency and the bandwidth of the narrow-band mode are 317~MHz and 33~MHz, respectively, and those of the wide-band mode are 400~MHz and 200~MHz, respectively. \rev{The field of view and the angular resolution are 75~arcmin and 8.3~arcsec, respectively, both in diameter at 400~MHz.}

The observations were carried out in Cycle 39 and were split into two separate observations on 13--14 January 2021 (day~1) and 24--25 February 2021 (day~2) in International Atomic Time (TAI). The observing time was 5 hours and 15 minutes on day~1, and 4 hours and 16 minutes on day~2. In each day, we observed a flux density, bandpass, and polarization calibrator, 3C286, for 10 minutes at the beginning and the end of the observation, and observed a phase calibrator, 1349-393, for 5 minutes every 25 minutes. Thus, the target (CIZA1359) on-source time was 337 minutes in total, 196 minutes on day 1 and 141 minutes on day 2. According to the observation log, two 45~meter-diameter antennas, C03 and C11, were not used on day~2. Therefore, only day~1 data were used in this study, as day~2 data tend to be a little noisy.

\subsection{The Data Reduction}

\begin{table*}[htbp]
\tbl{The radio maps in this paper.}{%
\begin{tabular}{lllllll} \hline \hline
Label & Frequency & BW & R.M.S. & Beam size & Beam PA & Figure \\
Unit & MHz & MHz & mJy~beam$^{-1}$ & asec $\times$ asec & degree &\\ \hline
Narrow-band & 317 & 33 & $8.2 \times 10^{-2} $ & 21.0 $\times$ 6.9 & -0.1 & - \\
Wide-band & 400 & 200 & $3.7 \times 10^{-2} $ & 14.8 $\times$ 5.2 & -6.2 & Figure \ref{fig:radio image2}\\
Sub-band 01 & 317 & 33 & $8.6 \times 10^{-2} $ & 22.8 $\times$ 5.7 & 1.6 & Figure \ref{fig:spec}\\
Sub-band 02 & 350 & 33 & $1.7 \times 10^{-1} $ & 15.1 $\times$ 5.0 & -0.5 & Figure \ref{fig:spec}\\
Sub-band 03 & 385 & 33 & $8.1 \times 10^{-2} $ & 17.8 $\times$ 5.6 & -0.3 & Figure \ref{fig:spec}\\
Sub-band 04 & 417 & 33 & $5.7 \times 10^{-2} $ & 16.2 $\times$ 5.4 & 1.6 & Figure \ref{fig:spec}\\
Sub-band 05 & 450 & 33 & $5.0 \times 10^{-2} $ & 14.7 $\times$ 4.9 & -1.1 & Figure \ref{fig:spec}\\
Sub-band 06 & 481 & 33 & $1.1 \times 10^{-1} $ & 12.7 $\times$ 4.6 & -1.4 & Figure \ref{fig:spec}\\
Smoothed Wide-band & 400 & 200 & $1.0 \times 10^{-1} $ & 25 $\times$ 25 & 0.0 & Figure \ref{fig:radio image2}, \ref{fig:radio image3}, \ref{fig:spec}, \ref{fig:source U} \\ \hline
\end{tabular}} \label{tab:2}
\end{table*}
The data were analysed using the SPAM (Source Peeling and Atmospheric Modeling; \cite{2014ASInC..13..469I}) which is based on the AIPS (Astronomical Image Processing System), produced and maintained by NRAO. The SPAM employed the AIPS 31DEC13 and was controlled by python 2.7. There is a bright compact source of about $2 \times 10^{3}$~mJy in the field of view, and the beam pattern of this bright source is clearly visible in the dirty image, showing that this source contaminates the image by causing strong sidelobes. Therefore, we applied the SPAM's Direction-Dependent Calibration (DDC; \cite{2017A&A...598A..78I}) to improve the dynamic range of the final image. Self-calibration is also applied to the Direction-Independent Calibration (DIC) before the DDC. The CLEAN algorithm is used for imaging in the SPAM.

In the analysis of the narrow-band data, we aim to create a catalog of sources for use in the DDC for the wide-band data. To select peeling sources, a list of radio sources in the field-of-view was compiled using a source catalog from the TIFR GMRT Sky Survey (TGSS) in this analysis. Imaging in the SPAM was performed with the Briggs robustness parameter of -1.0. The Python Blob Detector and Source Finder (PyBDSF; \cite{2015ascl.soft02007M}) was used to catalog the compact sources in the field-of-view.

In the analysis of the wide-band data, the data were split into six subbands and the DDC was applied to each subband data using the source catalog obtained from the analysis of the narrow-band data. The final output from the SPAM was the outlier-removed uv data. After that, a full-band radio image was derived by combining the uv data of subbands using WSClean (w-stacking clean; \cite{2014MNRAS.444..606O}). In the imaging with WSClean, we first employ {\it uniform} weighting of the robustness to identify compact sources from the data. The modeling and subtraction of the compact sources can reduce sidelobes of them, particularly from bright ones. Next, a multi-scale CLEAN \citep{2008ISTSP...2..793C} was performed with the robustness closer to {\it natural} weighting to derive the diffuse emission from the data.

The primary beam effect was corrected by using the AIPS task {\it pbcor}. 
\rev{
The function used to the applied primary beam model is $f(x) = 1.0 - \frac{2.939x}{10^{3}} + \frac{33.312x^2}{10^{7}} - \frac{16.659x^3}{10^{10}} + \frac{3.066x^4}{10^{13}}$, where $x$ is distance parameter (see AIPS Cookbook for details).
}
AIPS was also used to edit the data at each frequency to the same pixel size and spatial resolution, if necessary.

When calculating the flux density $F_\nu$, \rev{we adopted an empirical 10\% error ($\sigma_{{\rm abs}} = 0.1$) of the absolute flux density (see also Section 3.1 according to the DDC flux decay).} For a diffuse source, the flux density error, $\sigma_{F_{\nu}}$, was given by $\sigma_{F_{\nu}} = \sqrt{\left( \sigma_{rms} \sqrt{N_{\rm b}} \right) ^2 + \left( \sigma_{{\rm abs}} F_\nu \right) ^2 }$, where $\sigma_{{\rm rms}}$ is the rms noise in an image, and $N_{\rm b}$ is the number of beams in the diffuse source (e.g., \cite{2022MNRAS.tmp.1605K}).

\subsection{Other Data}

In this study, X-ray data from Suzaku are used to confirm the spatial correlation with the ICM \citep{2015PASJ...67...71K}. In addition, the ICM temperature inferred from the XMM-Newton data was used as an indicator of the shock region \citep{2022arXiv221002145O}. The MeerKAT Galaxy Cluster Legacy Survey Data Release 1 (MGCLS DR1; \cite{2022A&A...657A..56K}) were also combined to determine the spectral index. The used MGCLS data were {\it Enhanced imaging products}, which was corrected the primary beam effect.

\section{Results}

\subsection{Total Intensity Maps}


\begin{figure*}[tbp]
 \begin{center}
  \includegraphics[width=0.9\linewidth]{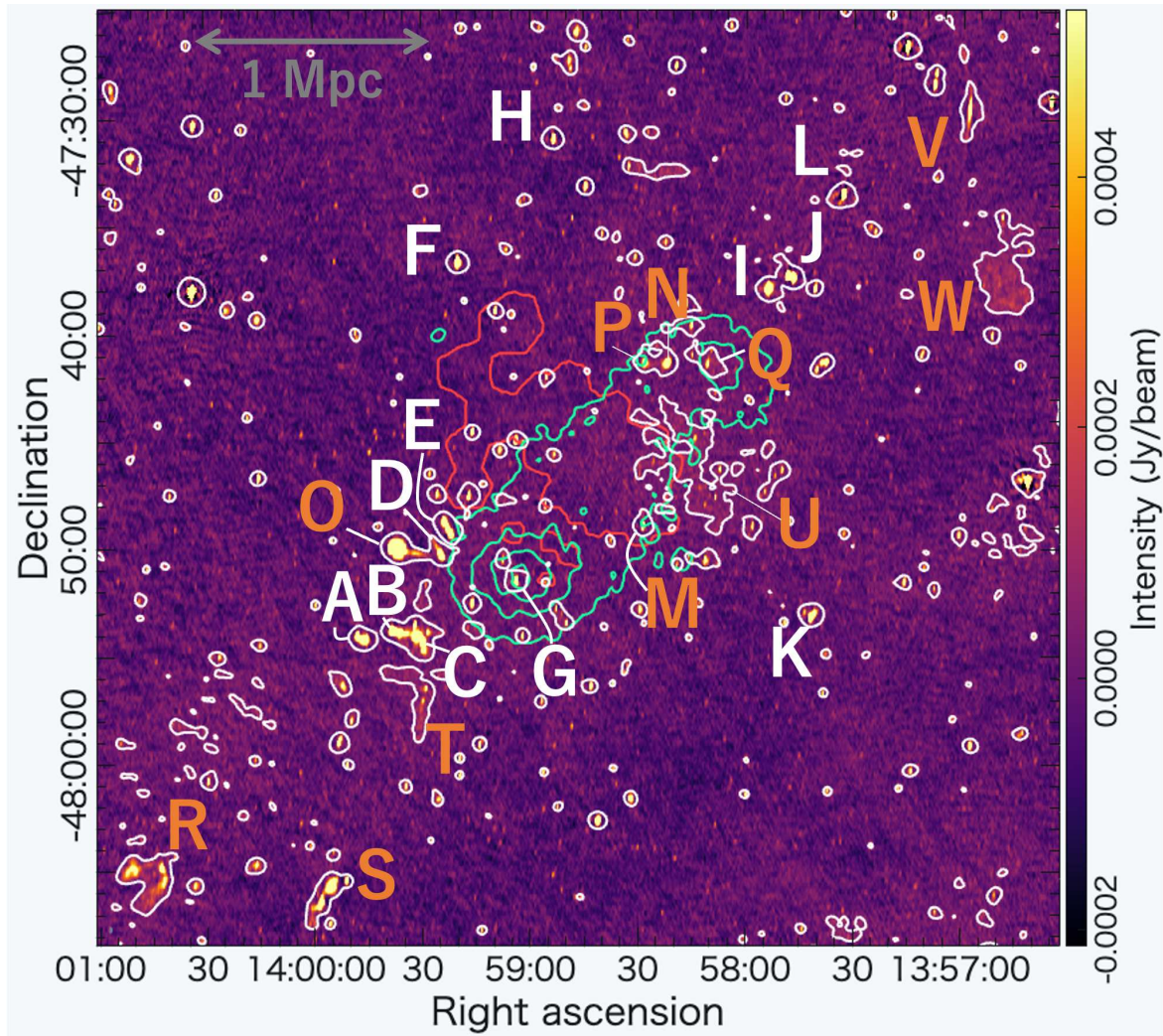} 
 \end{center}
\caption{Wide-band image of CIZA1359. The background image is the total intensity at 400~MHz with the bandwidth of 200~MHz. The beam pattern with the size, $14.''8 \times 5.''2 $, is shown in the gray filled elliptical at the lower left corner of the figure. The white contours show the intensity level of $0.4$~mJy~beam$^{-1}$ which corresponds to $4\sigma_{rms}$ with the smoothed 25~arcsec resolution. The green contours show the X-ray surface brightness distribution of Suzaku \citep{2015PASJ...67...71K} in arbitral units at intervals of 1.81, 3.64, 5.46 and 7.28. The red contour shows the region with a temperature of 6 keV or above inferred from the recent X-ray observations \citep{2022arXiv221002145O}. The diffuse radio sources, including relic candidate, are labeled with the A to W.}\label{fig:radio image2}
\end{figure*}

\begin{table*}[htbp]
\tbl{Imaging parameters performed to focus on the diffuse emission.}{%
\begin{tabular}{clllll} \hline \hline
Weighting & R.M.S. & Beam size & Beam PA & Taper & Figure \\
Unit & mJy\~beam$^{-1}$ & asec $\times$ asec & degree & arcsec & \\  \hline 
uniform & $0.18$ & $10.5 \times 3.2$ & $-2.4$ & 22.5 & - \\
briggs -1 & $1.23$ & $25.2 \times 21.2$ & 24.7& 22.5 & - \\
briggs 0 & $1.56$ & $36.8 \times 22.7$ & 29.6& 22.5 & Figure \ref{fig:source U} (a) \\
briggs 1 & $6.24$ & $296.3 \times 40.6$ & 24.7& 22.5 & - \\ \hline
 \end{tabular}}\label{tab:4}
\end{table*}

We derived the total intensity map of the narrow-band data at 317~MHz. The DDC was applied to the imaging and the robustness parameter of $-1.0$ was set. The rms noise level of the map is $8.2 \times 10^{-2}$~mJy~beam$^{-1}$ with the DDC (table \ref{tab:2}), while that is $1.73$~mJy~beam$^{-1}$ with the IDC. Therefore, the DDC improved the sensitivity and dynamic range by more than one order of magnitude. We found and cataloged 423 radio sources by the PyBDSF. The brightest source is PMN J1401-4733 at the north-east of the field of view, with the total flux density of $2.01 \times 10^{3}$~mJy and a peak intensity of $1.41 \times 10^{3}$~mJy~beam$^{-1}$. Thus, the achieved image dynamic range is 17,195.

Next, we derived the total intensity map of the wide-band data at 400~MHz. We used the above source catalog as a prier sky model for the DDC, then obtained the rms noise level of the map, $3.7 \times 10^{-2}$~mJy~beam$^{-1}$ (table \ref{tab:2}), \rev{or the achieved image dynamic range of 38,108} for the robustness parameter of $-1.0$. The wide-band image of CIZA1359 is shown in figure \ref{fig:radio image2}. The background colour is the radio intensity distribution, where the white contour indicates the $0.4$~mJy~beam$^{-1}$ level (4~$\sigma_{\rm rms}$) for the uGMRT smoothed at $25''$.  The alphabetic labels indicate 23 distinguished sources. The white labels from A to L are known sources, while the orange labels from M to W are newly-detected\footnote{After we submitted this paper, \citet{2022MNRAS.tmp.1605K} reported Sources from M to Q.} extended sources. 

\begin{figure*}[tbp]
 \begin{center}
  \includegraphics[width=0.45\linewidth]{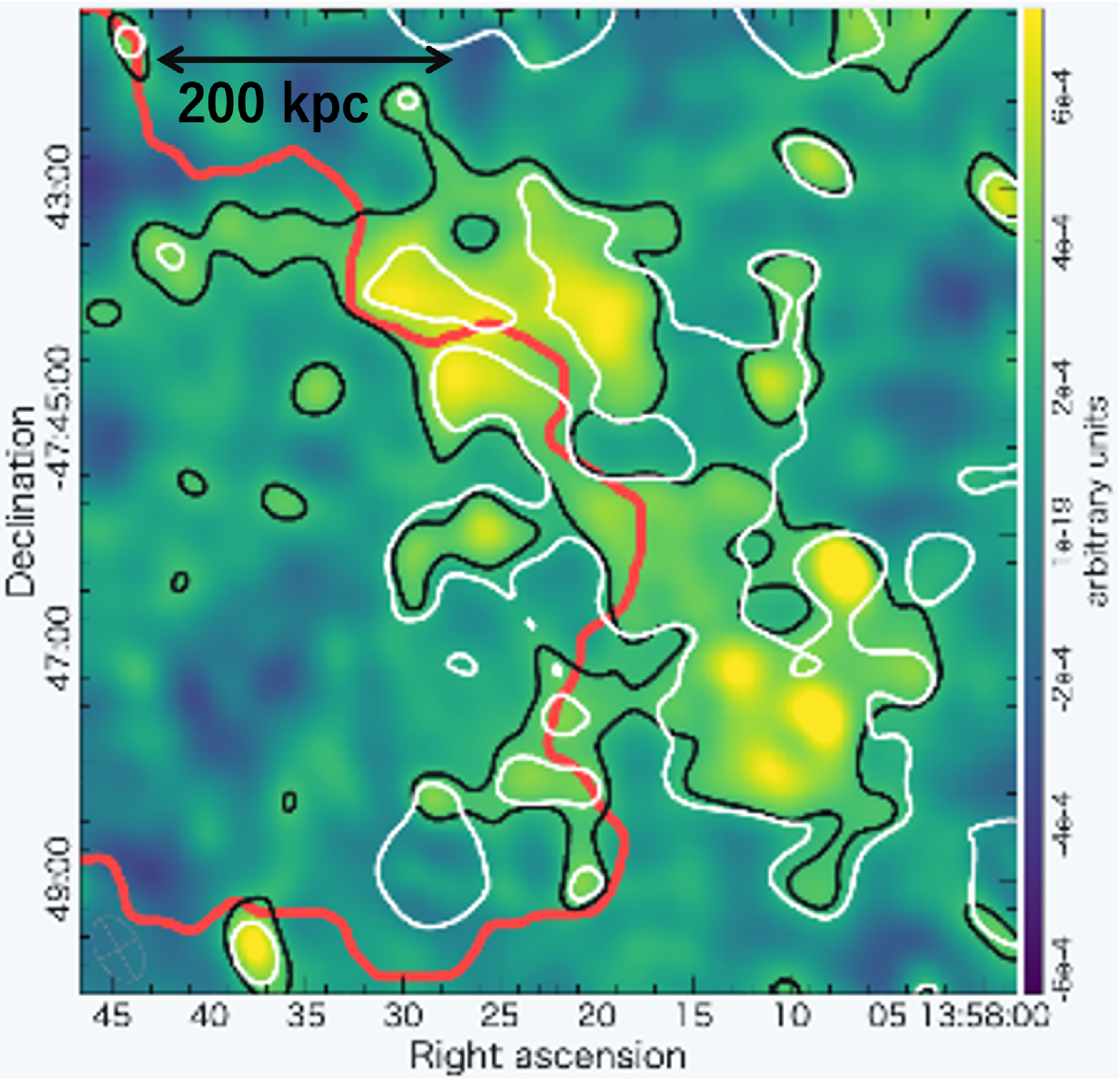} 
  \includegraphics[width=0.45\linewidth]{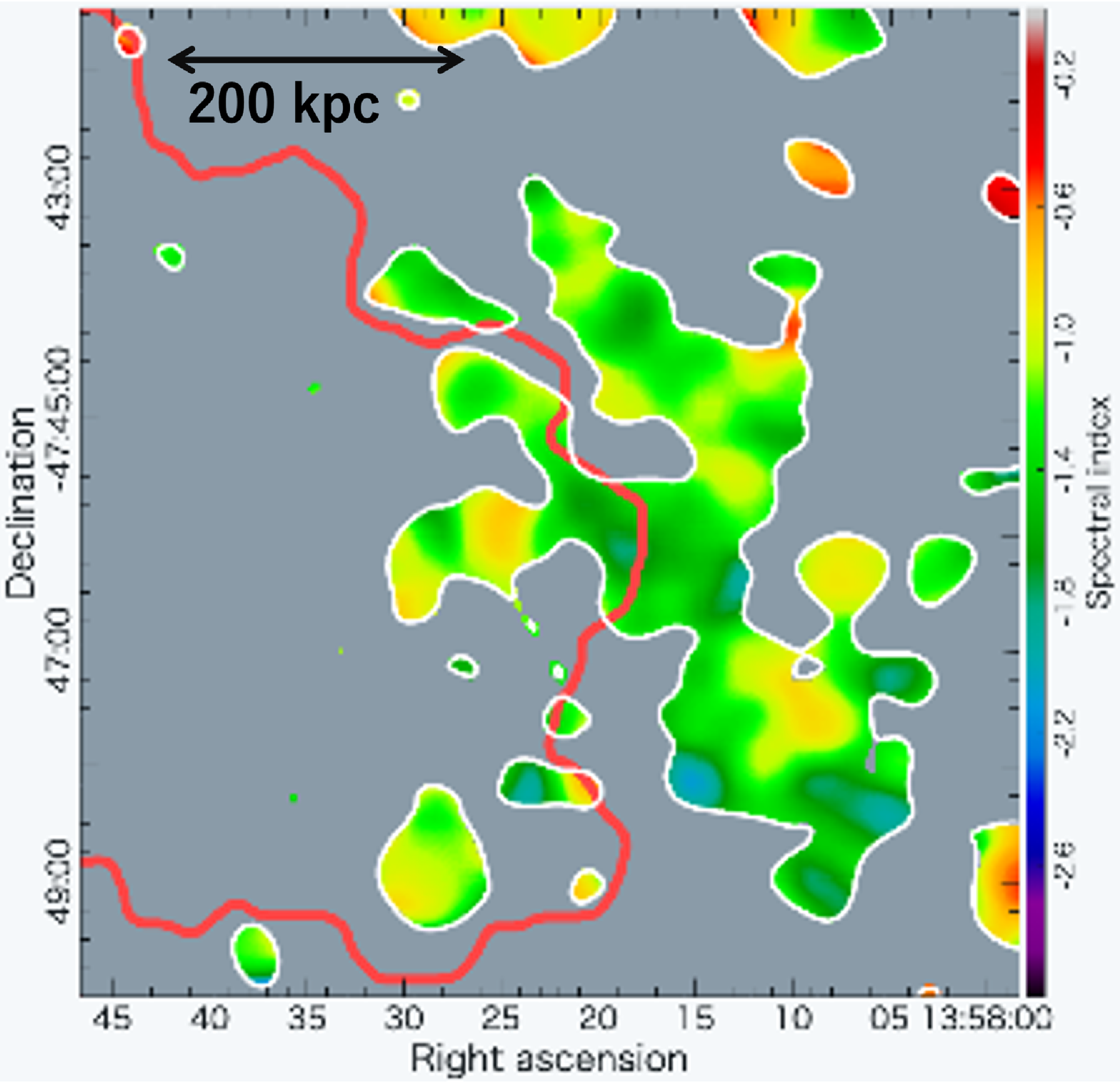} 
 \end{center}
\caption{Radio and spectral maps of Source U. White and red contours are the same as in figure \ref{fig:radio image2}. {\it Left (a):} The background color shows the point source subtracted intensity. 
\rev
{This unit is equivalent to Jy/beam. But the point source is subtracted with a Gaussian fit, we expressed an "arbitrary unit" in the sense that it is different from the color unit in figure \ref{fig:radio image2}.
}
The black contours have a point source subtracted intensity of $3.0 \times 10^{-4}$, which corresponds 3$\sigma_{rms}$. {\it Right (b):} It is a spectral index map of source U. }\label{fig:source U}
\end{figure*}

\rev{One of the known issues on the DDC analysis is that a lot of DDC solutions result in global decay of the fluxes across the field (e.g., \cite{2016MNRAS.463.4317P}). To assess this decay, we checked the visibility amplitude with respect to the DDC and DIC results. The maximum angular scale of source U, which is the largest feature among the structures detected in this paper, is 6 arcminutes, which is about 0.6 kilo wavelength at 300 MHz. The medians of the visibility amplitude over a range of 0.6 kilo wavelength are 1.168~Jy for the DDC and 1.247~Jy for the DIC, respectively. Therefore, we measured an offset of about 0.94 in the amplitude ratio around the angular scale of source U. The error can be compared to an empirical 10 \% error of the absolute flux we display in this paper. The Flux accuracy checks are summarized in Appendix A.}

\rev{To facilitate our discussion, a compact-source-subtracted image for Source U was derived. The compact sources were subtracted from the image by fitting with a point-source model with a Gaussian function using PyBDSF. We explored the robustness closer to {\it natural} weighting to image the diffuse emission. Table \ref{tab:4} summarize the noise level and resolution from the CLEAN with different weightings; we employed the robustness of $0.0$. The results are shown in figure \ref{fig:source U} (a). The hot ICM region is shown as red contours, which would indicate the approximate location of the merger shock front. The white contours are the same as that in figure \ref{fig:radio image2}. The black contours corresponds to 3$\sigma_{rms}$ in the compact-source-subtracted image which is also shown in the background color.}

\subsection{Spectral Index Map}


\begin{figure*}[tbp]
 \begin{center}
 \includegraphics[width=0.3\linewidth]{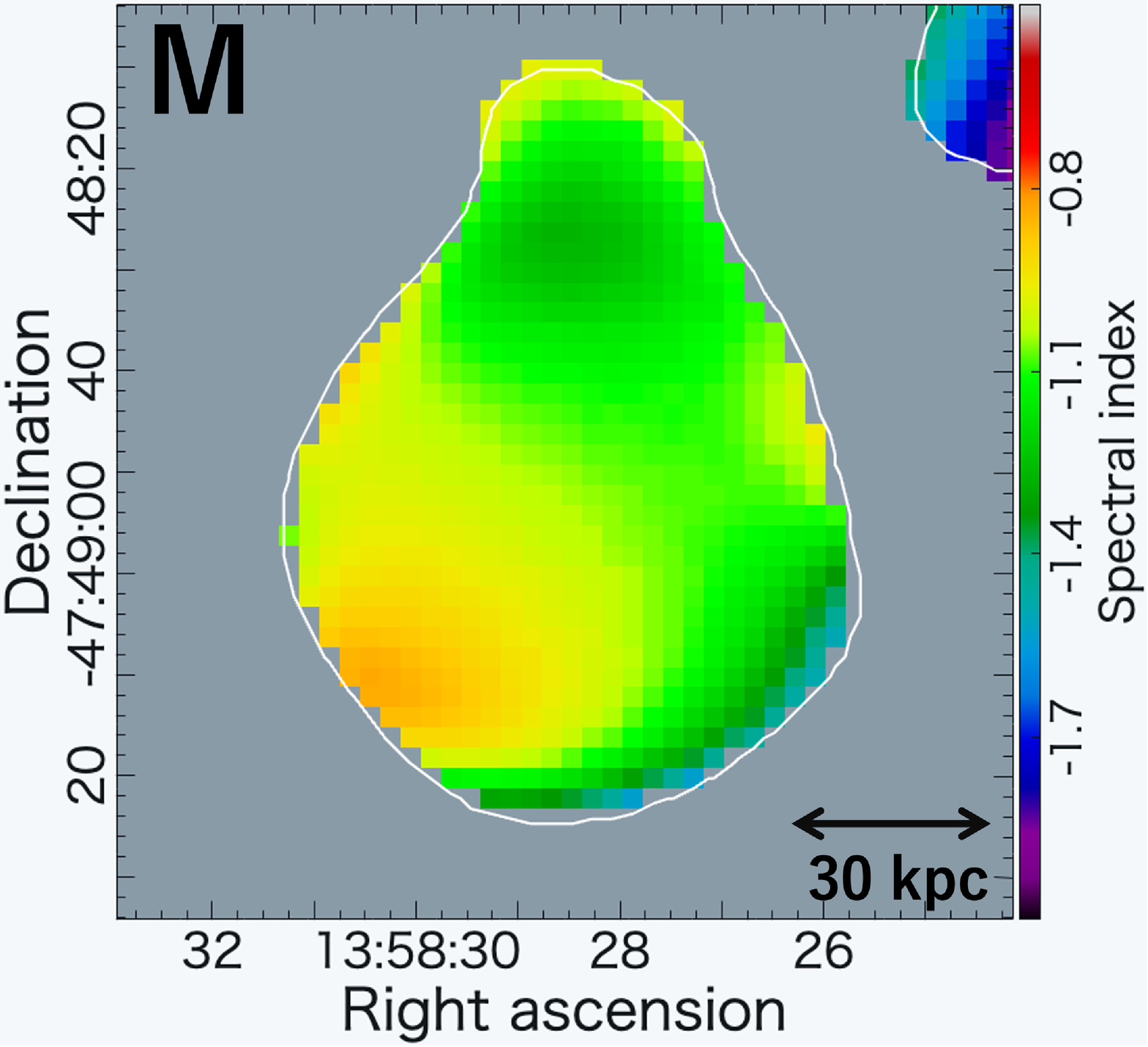} 
 \includegraphics[width=0.3\linewidth]{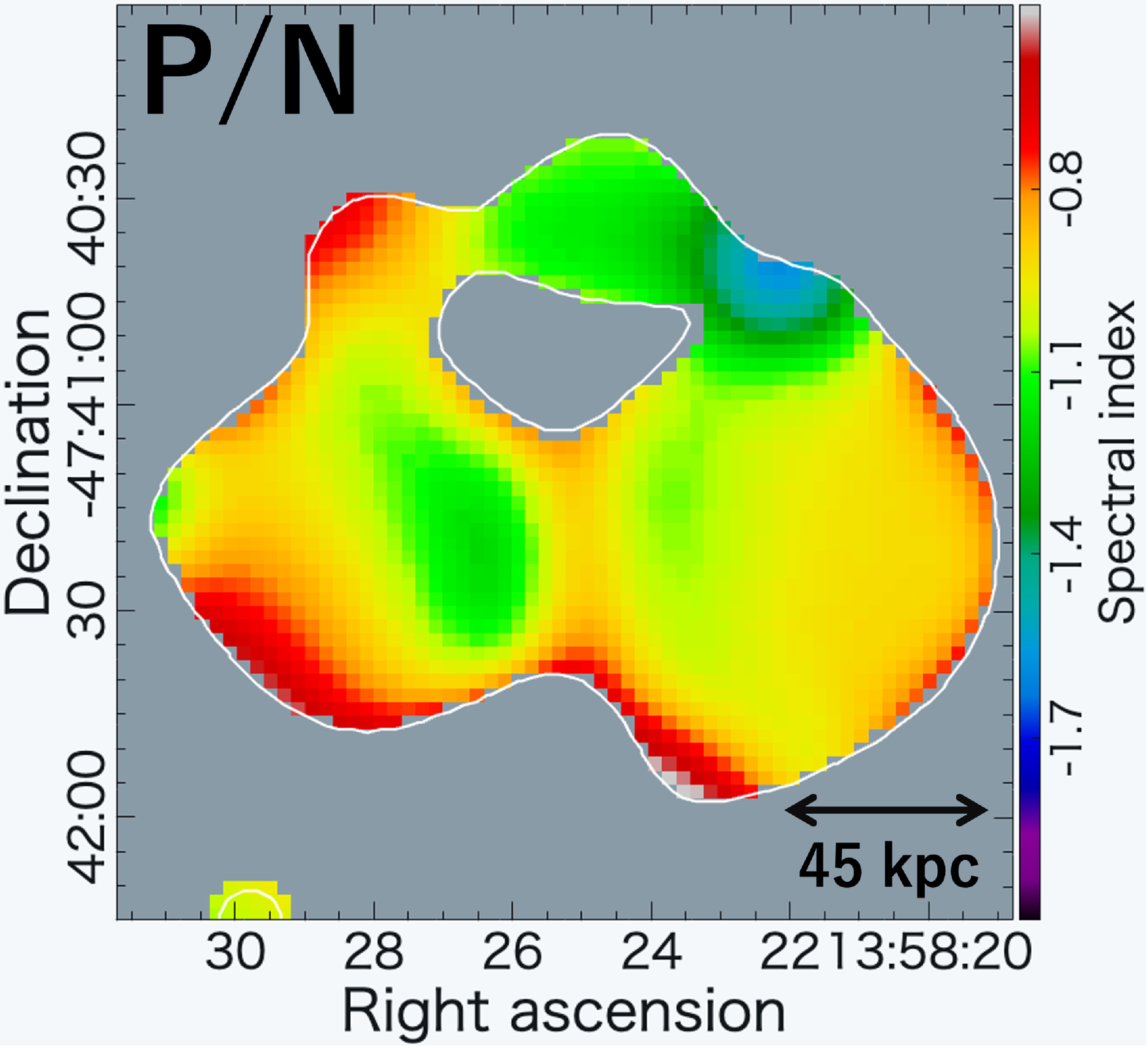} 
 \includegraphics[width=0.3\linewidth]{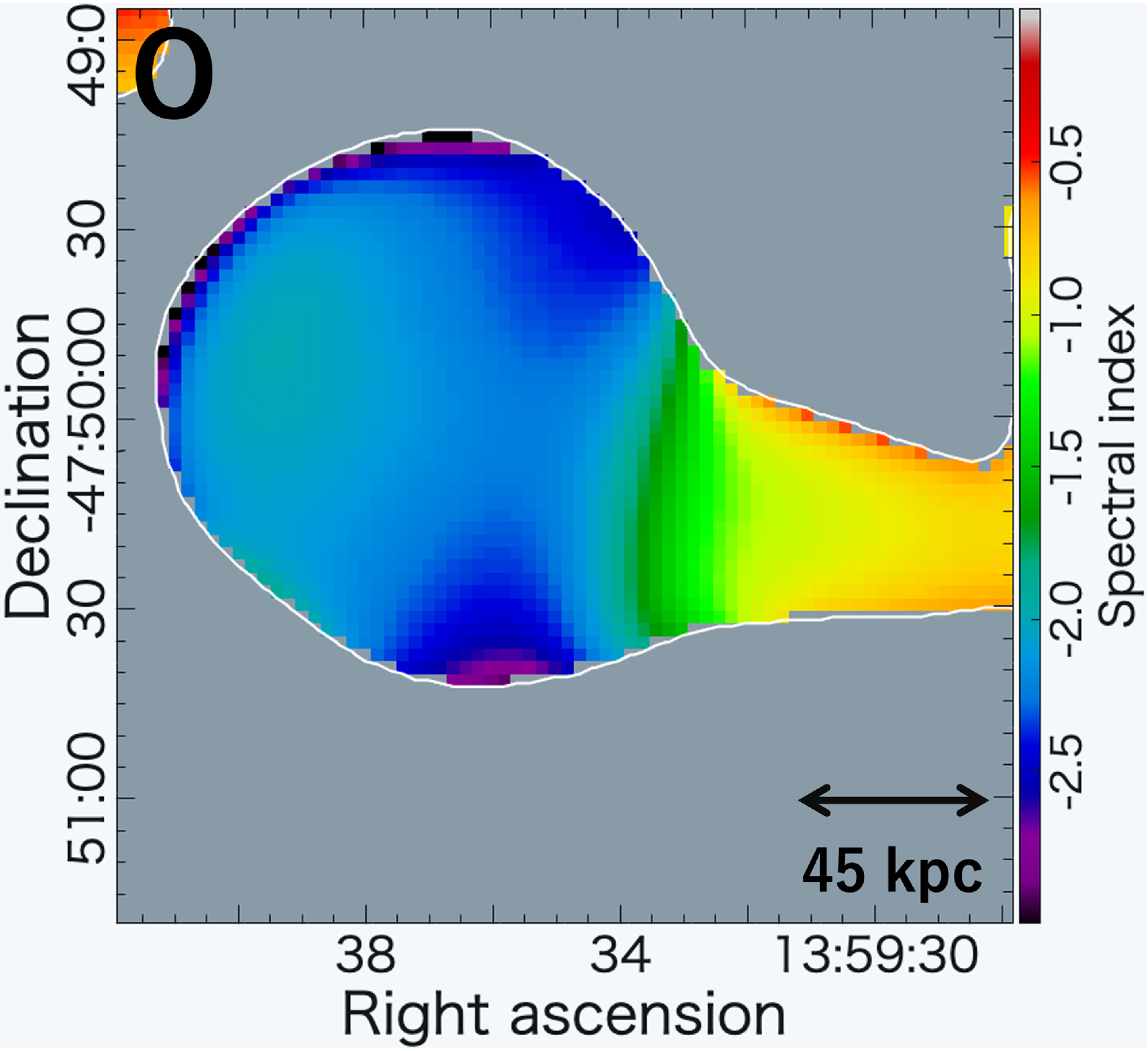} 
 \includegraphics[width=0.3\linewidth]{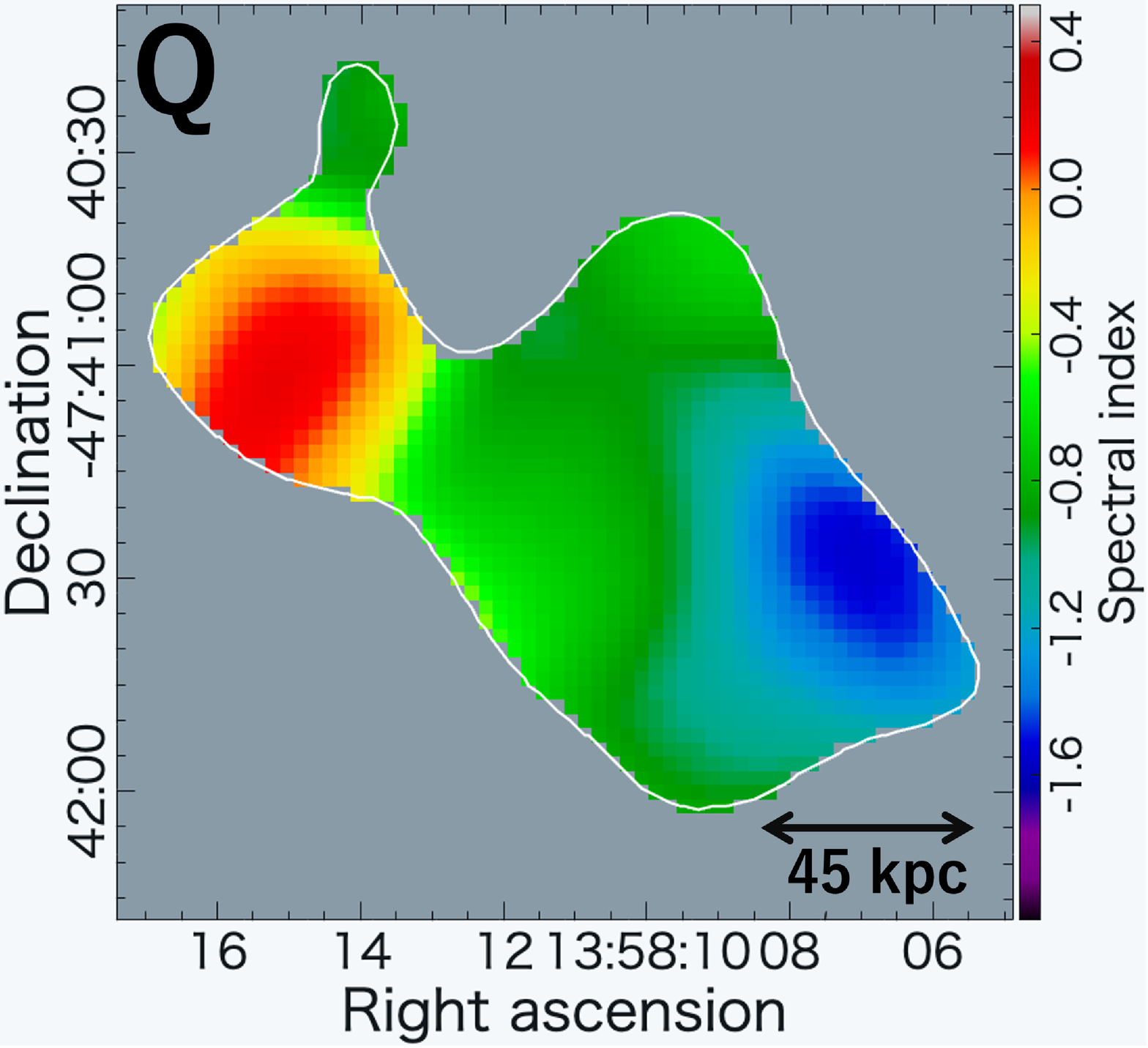} 
 \includegraphics[width=0.3\linewidth]{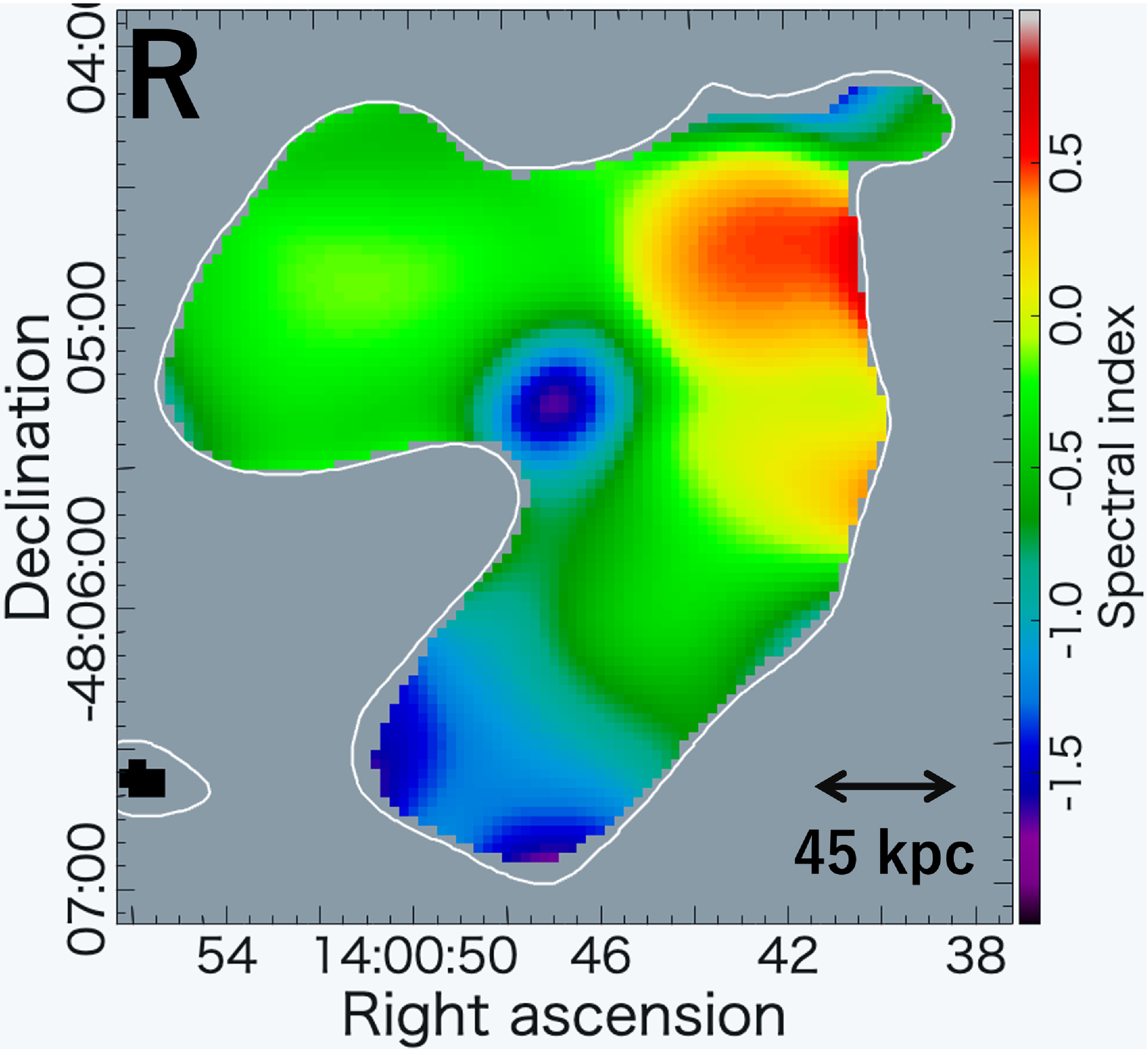} 
 \includegraphics[width=0.3\linewidth]{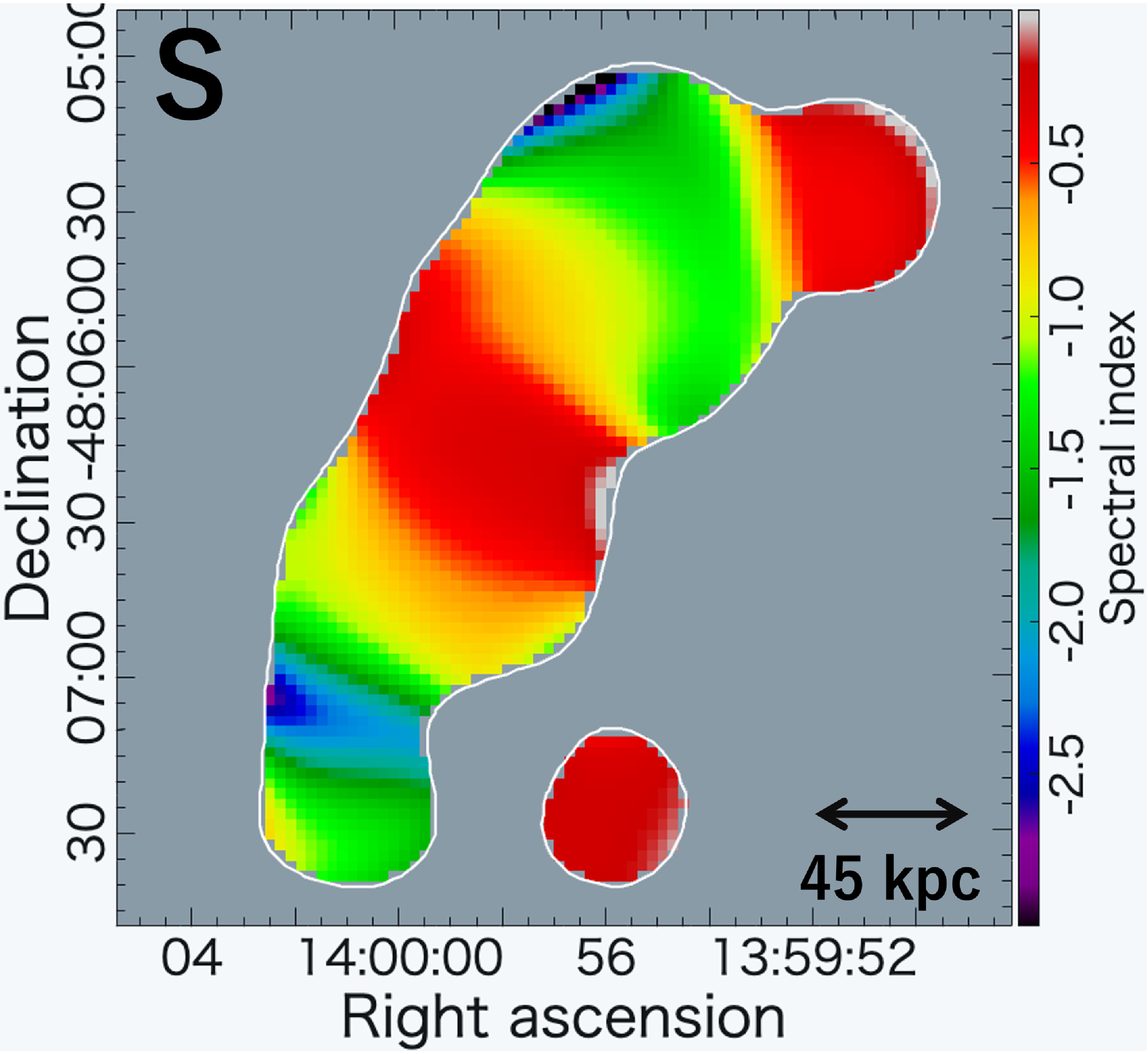} 
 \includegraphics[width=0.3\linewidth]{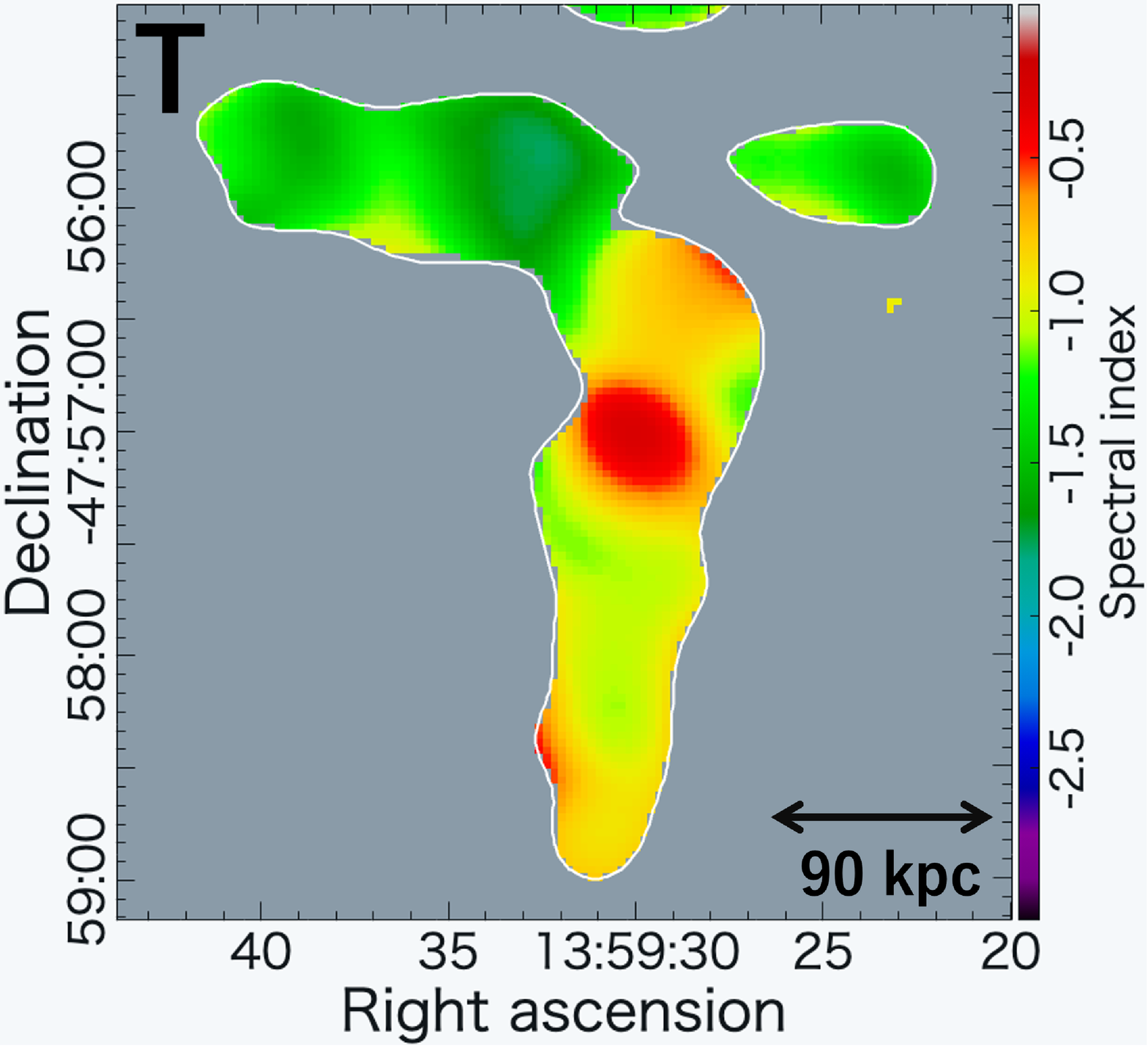} 
 \includegraphics[width=0.3\linewidth]{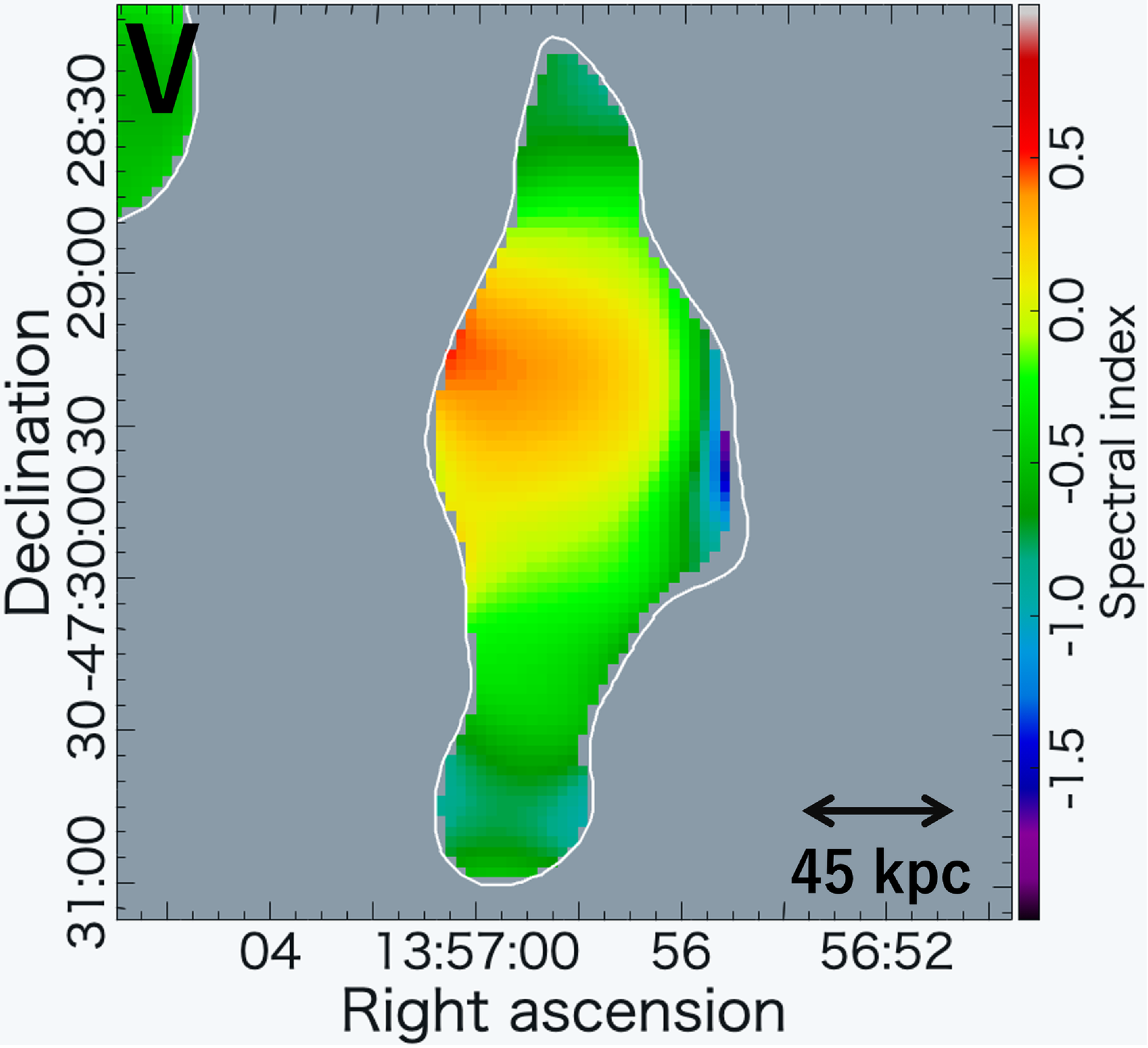} 
 \includegraphics[width=0.3\linewidth]{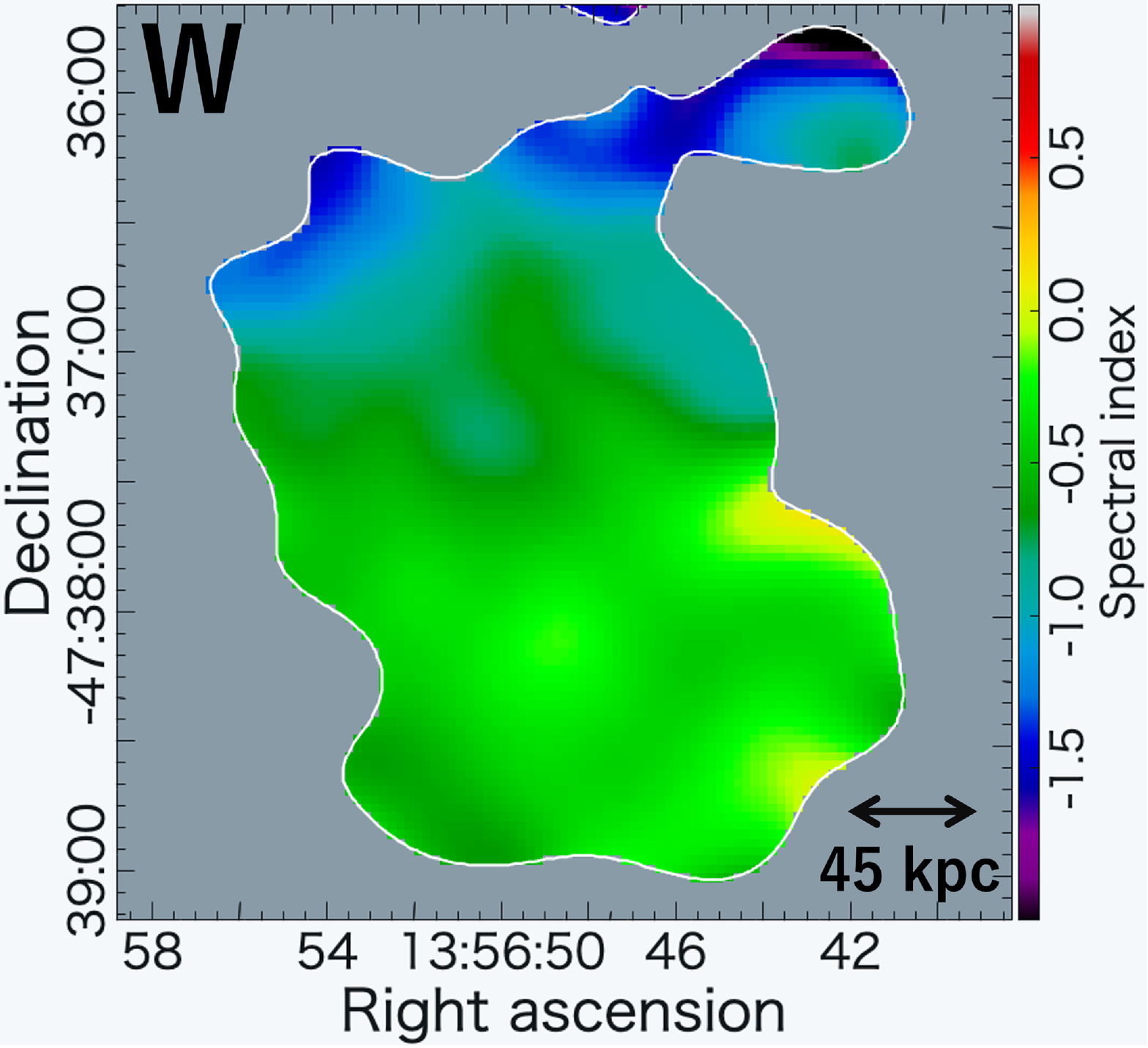} 
 \end{center}
\caption{Spectral index maps of newly detected extended sources in CIZA1359. The spectral index $\alpha$ is calculated using $F_\nu \propto \nu^{\alpha}$. The white contours are the same as figure \ref{fig:radio image2}. Note that the color range is different for each.}\label{fig:radio image3}
\end{figure*}

We convolved all images to a 25-arcsec square beam using the AIPS task {\it convl}, where the pixel size and the number of pixels were fixed using the AIPS task {\it hgeom}. We then adopted the least-square fit to derive the best-fit spectral index assuming a power-law and calculated the index pixel by pixel. We performed the fitting in a linear space to account for the negative flux value caused by the noise. To derive the spectral index, we added the MGCLS DR1 data (see Section 2.3) to our uGMRT data. Details on the calculation of the spectral index are summarized in Appendix B. \rev{Figure \ref{fig:source U} (b) and} figure \ref{fig:radio image3} show the spectral index maps of \rev{newly detected} extended sources in CIZA1359. The background colour indicates the spectral index, $\alpha$, such that $F_\nu \propto \nu^{\alpha}$, and the white contour is the same as that in figure \ref{fig:radio image2}. 

\rev{We also attempted to calculate in-band spectral indices for bright compact sources using subband images, for each GMRT and MeerKAT data. We obtained approximately the same index no matter which data is used for fitting for the sources such as Source A.  On the other hand, because faint sources including many diffuse-emission features have a low SN in each pixel, the subband spectral fitting results in the spectral index close to the slope of the noise floor. Therefore, for the faint sources we used the combined data in each band, where the center frequencies are 400 MHz and 1280 MHz, to derive the spectral index, $\alpha_{400-1280}$.}

In addition to the spectral index of each pixel, we derived the mean spectral index using the total flux densities shown in columns 4 and 5 of table \ref{tab:3}. The total flux density was estimated by assuming the size of each source. The resultant mean spectral index is listed in column 6 of table \ref{tab:3}. Here, the spectral index error was derived from the error propagation equation ($\sqrt{(F_{\nu 1}~{\rm log}(\nu_1 / \nu_2 ) )^{-2} \sigma_{F_{\nu 1}}^2+(-F_{\nu 2}~{\rm log}(\nu_1 / \nu_2 ))^{-2} \sigma_{F_{\nu 2}}^2}$).

\subsection{Source Catalog}

\begin{table*}[tbp]
\tbl{Radio source catalog of the CIZA1359 field. Columns are: (1) Source name. (2) Average intensity in mJy~beam$^{-1}$, which adopts the white contours of figure \ref{fig:radio image2} as the size of the source. The error can be used $3.7 \times 10^{-2}$~mJy~beam$^{-1}$. (3) Peak intensity in mJy~beam$^{-1}$, size of the source is the same as column 2. The error is used a rms noise around the source in the point source subtracted image. (4) Integrated flux which is not subtracting point sources in mJy at 400~MHz, size of the source is the same as column 2. (5) Same as column 4 but frequency is 1280~MHz. (6) Spectral index, which is calculated using the fluxes in columns 4 and 5. (7) Corresponding galaxies nearby in sky-plane.
}{%
\begin{tabular}{ccccccc} \hline \hline
Label & Mean intensity & Peak Flux & \multicolumn{2}{c}{Flux} & $\alpha_{400-1280}$ & Identification \\
 & 400~MHz$^{*1}$  & 400~MHz$^{*1}$  & 400~MHz$^{*1}$  & 1280~MHz$^{*2}$ &  &  \\ 
(1) & (2)  & (3)  & (4)  & (5)  & (6) & (7)  \\ \hline
A & 4.74 & $ 17.85 \pm 0.21 $ & $ 25.90 \pm 2.59 $ & $ 13.10 \pm 1.32 $ & $ - 0.59 \pm 0.28$ & - \\
B & 4.86 &  $ 17.41 \pm 0.11 $ & $ 29.62 \pm 2.97 $ & $ 13.50 \pm 1.37 $ & $ - 0.68 \pm 0.28 $ & 2MASX J13593870-4753472 \\
C & 3.72 &  $ 17.54 \pm 0.12 $ & $ 43.54 \pm 4.36 $ & $ 17.49 \pm 1.77 $ & $ - 0.78 \pm 0.28 $ & 2MASX J13593065-4754053 \\
D & 2.47 &  $ 10.47 \pm 0.16 $ & $ 14.04 \pm 1.42 $ & $ 6.35 \pm 0.67 $ & $ - 0.68 \pm 0.29 $ & 2MASX J13592419-4750253  \\
E & 3.26 &  $ 11.29 \pm 0.23 $ & $ 21.98 \pm 2.21 $ & $ 10.84 \pm 1.11 $ & $ - 0.61 \pm 0.28 $ & 2MASX J13592518-4749333 \\
F & 2.56 &  $ 9.28 \pm 0.08 $ & $ 20.30 \pm 2.04 $ & $ 9.58 \pm 0.97 $ & $ - 0.65 \pm 0.28 $ & - \\
G & 1.18 &  $ 4.63 \pm 0.10 $ & $ 10.93 \pm 1.11 $ & $ 4.88 \pm 0.52 $ & $ - 0.69 \pm 0.29 $ & 2MASX J13590381-4751311 \\
H & 2.48 &  $ 9.63 \pm 0.09 $ & $ 15.31 \pm 1.54 $ & $ 9.37 \pm 0.95 $ & $ - 0.42 \pm 0.28 $ & -  \\
I & 2.91 &  $ 13.71 \pm 0.14 $ & $ 31.22 \pm 3.13 $ & $ 15.12 \pm 1.52 $ & $ - 0.62 \pm 0.28 $ & 2MASX J13575383-4737543  \\
J & 5.79 &  $ 28.21 \pm 0.26 $ & $ 38.69 \pm 3.88 $ & $ 17.56 \pm 1.77 $ & $ - 0.68 \pm 0.28 $ & -\\
K & 3.12 &  $ 11.54 \pm 0.18 $ & $ 15.18 \pm 1.53 $ & $ 6.16 \pm 0.64 $ & $ - 0.78 \pm 0.29 $ & -  \\
L & 5.47 &  $ 30.51 \pm 0.07 $ & $ 32.66 \pm 3.27 $ & $ 13.51 \pm 1.37 $ & $ - 0.76 \pm 0.28 $ & -\\
M & 2.20 &  $ 8.77 \pm 0.13 $ & $ 9.11 \pm 0.93 $ & $ 2.73 \pm 0.32 $ & $ - 1.04 \pm 0.31 $ & - \\
N & 1.81 &  $ 8.20 \pm 0.16 $ & $ 11.13 \pm 1.13 $ & $ 3.63 \pm 0.42 $ & $ - 0.96 \pm 0.30 $ & 1RXS J135821.7-474126 \\
O & 4.70 &  $ 19.57 \pm 0.63 $ & $ 50.57 \pm 5.06 $ & $ 4.87 \pm 0.56 $ & $ - 2.01 \pm 0.30 $ & - \\
P & 1.30 &  $ 4.12 \pm 0.14 $ & $ 5.55 \pm 0.58 $ & $ 1.83 \pm 0.25 $ & $ - 0.95 \pm 0.34 $ & - \\
Q & 0.72 &  $ 1.99 \pm 0.19 $ & $ 5.52 \pm 0.60 $ & $ 2.59 \pm 0.35 $ & $ - 0.65 \pm 0.34 $ & 2MASX J13581085-4741243 \\
R & 1.39 &  $ 5.60 \pm 0.42 $ & $ 31.06 \pm 3.13 $ & $ 27.22 \pm 2.75 $ & $ - 0.11 \pm 0.28 $ & 2MASX J14004272-4804474 \\
S & 2.41 &  $ 9.79 \pm 0.29 $ & $ 31.32 \pm 3.15 $ & $ 13.29 \pm 1.36 $ & $ - 0.74 \pm 0.28 $ & - \\
T & 0.65 &  $ 2.07 \pm 0.32 $ & $ 10.06 \pm 1.06 $ & $ 3.43 \pm 0.48 $ & $ - 0.93 \pm 0.35 $ & 2MASX J13592976-4757043 \\
U & 0.54 &  $ 1.29 \pm 0.15 $ & $ 28.93 \pm 2.96 $ & $ 7.03 \pm 0.94 $ & $ - 1.22 \pm 0.33 $ & 2MASX J13580947-4745213\\
&&&&&&\begin{tabular}{c}
2MASX J13581294-4748183\\
6dFGS gJ135839.1-474723\\
\end{tabular} \\
V & 1.57 &  $ 7.34 \pm 0.34 $ & $ 12.75 \pm 1.30 $ & $ 14.39 \pm 1.46 $ & $ 0.10 \pm 0.28 $ & 2MASX J13565832-4729231  \\
W & 0.74 &  $ 1.24 \pm 0.20 $ & $ 21.34 \pm 2.18 $ & $ 11.60 \pm 1.25 $ & $ - 0.52 \pm 0.29 $ & -  \\
\hline \end{tabular}}\label{tab:3}
\begin{tabnote}
${*1}$: This work; ${*2}$: MGCLS \citep{2022A&A...657A..56K} 
\end{tabnote}
\end{table*}

We found 23 distinguished radio features in the image. We labeled them from Sources A to W (figure \ref{fig:radio image2}). The parameters for each source are summarised in table \ref{tab:3}. 

Sources from A to L were reported in the previous ATCA observation \citep{2018PASJ...70...53A}, where several sources are closely concentrated on the south-western rim of the southern subcluster of CIZA1359. The spectral indices are comparable to those estimated with ATCA and those of typical AGN. We found that some of the sources were clearly larger in spatial scale than the synthesized beam and were resolved into multiple components.  Source G is located near the center of the south subcluster and has an apparent size of about $25'' \times  22''$ (=38~kpc x 33~kpc) at the 3$\sigma_{rms}$ signal to noise level. It is one order of magnitude smaller than the typical size of mini-halos ($\sim 500$~kpc), so that it is more likely to be an AGN radio lobe. Source G is cataloged in SIMBAD as a galaxy 2MASX J13590381-4751311. It has the redshift of z=0.074, which is the consistent with the redshift of CIZA1359.

Sources from M to Q were reported in the previous GMRT observation \citep{2022MNRAS.tmp.1605K}. Similarly to Source G, Source Q is more likely to be an AGN, although it is located near the center of the north subcluster with a size of $60'' \times 30''$ (=90~kpc $\times $ 45~kpc). Source Q is cataloged in SIMBAD as a galaxy 2MASX J13581085-4741243. Its redshift, z=0.074, is consistent with the redshift of CIZA1359. Source O, which is unresolved in TGSS and SUMSS images, consists of a round structure and a narrow east-west linear structure. A high-resolution image such as MGCLS \rev{at 1.28 GHz} shows a head-tail galaxy-like structure, while there is no corresponding source in the ATCA image at \rev{2.1 GHz} \citep{2018PASJ...70...53A}. Indeed, a steep spectral index of $-2.01$ and the 400 MHz total flux density, $48.73 \pm 4.88$~mJy, predicts the 2~GHz flux density of 1.9~mJy (with a size of $90'' \times 90''$), which is almost the same as the sensitivity limit of the ATCA observation. Source O can be classified into an Ultra-Steep Spectrum (USS) source found in galaxy clusters \citep{2019A&A...622A..22M}, so that Source O may be a fossil plasma source. We could not find any corresponding source to Source O in SIMBAD.

Source R is about 2~Mpc away from the southern subcluster center to the south-east and is cataloged in SIMBAD as a galaxy 2MASX J14004272-4804474. We found that it has a radio structure like a head-tail galaxy. There is also a spectral index gradient, with aging from the head to the tail. It has the redshift of z=0.075, which is comparable to that of CIZA1359.

Source S has a FRII-like radio structure about 1.5~Mpc away from the southern subcluster center. SIMBAD cataloged a galaxy at z=0.054 as 2MASX J13595922-4805486 in the neighbourhood. However, 2MASX J13595922-4805486 may be associated with a faint radio structure seen at 11~arcsec away to the northeast from Source S.

Source T is located at the southeast of the CIZA1359 and has a head-tail galaxy-like radio structure. It has an elongated structure to the south and a faint structure to the east. There is no associated source within 30~arcsec in SIMBAD. At the peak flux position of the south component of Source T, there is a galaxy 2MASX J13592976-4757043 in SIMBAD. The redshift of 2MASX J13592976-4757043 is z=0.081, which is located at far side of CIZA1359.

Source U is a candidate of diffuse cluster emission. It is located at the south of the northern subcluster and in between the two subclusters. Four galaxies were found by SIMBAD within a 4~arcmin radius centred on source U; 2MASX J13580947-4745213 is located at the north-west of Source U with the redshift, z=0.078, 2MASX J13581294-4748183 is located at the south of Source U with the redshift, z=0.069, 6dFGS gJ135839.1-474723 is located at the east of Source U with the redshift, z=0.067, and LEDA 184317 is located at the north-east of Source U with an unknown redshift. Source U is the largest diffuse source in the uGMRT image (see figure \ref{fig:radio image2}). The signal to noise ratio of Source U is around 4~$\sigma_{{\rm rms}}$, where $\sigma_{{\rm rms}} = 0.10$~mJy~beam$^{-1}$. With $N_{{\rm b}} = 23.0$, the point source subtracted flux density is $24.04 \pm 2.48$~mJy, which significantly deviates from the null. We explore Source U in detail in Section 4.

Source V is cataloged in SIMBAD as a galaxy 2MASX J13565832-4729231, with the redshift, z=0.078, similar to that of CIZA1359. It has an elongated structure extending toward north-south. The extended structure appears to be connected to Source W; Source V is like a bipolar radio jets.

Source W is quadrangle in shape and is the second largest radio structure in the image. No corresponding sources were found in SIMBAD in the vicinity of this Source. The Sources V and W are similar to the structure known as the radio phoenix.

\section{Discussion}

We explore Source U in detail in the next section 4.1, followed by the discussion of its origin (likely a radio relic) supposing its detection in section 4.2.

\subsection{Source U: Diffuse Radio Structure Candidate}

\subsubsection{Location}

First, we focus on the spatial location of Source U. As described in Introduction, the recent X-ray observation found a pair of shock fronts in the linked region \citep{2022arXiv221002145O}; the north shock at the northern edge of the hot region, and the south shock at the southern edge of the hot region, where the hot region is shown as the red solid-line in figure \ref{fig:source U} (a). The pair shocks seem to be emerged from the interface of the subclusters and be propagating toward each subcluster core.

Such a merger shock has been considered as a site of cluster diffuse radio emission, based on an expectation that the shock accelerates cosmic-ray electrons emitting synchrotron radiation. Actually, the location of Source U is broadly consistent with that of the western part of the north shock indicated by the red contours in figure \ref{fig:source U} (a). In fact, the shape of Source U and the shock plane do not exactly coincide with each other. If Source U was excited by a shock wave, it flows down and ages at the downstream-side of the shock front. But it is located at the slightly upstream side. This would be interpreted due to the misalignment of the merger axis with the sky plane, i.e., a projection effect of the viewing angle.

\subsubsection{Structure}

The shock-associated diffuse radio emission is often seen in the late-stage merging clusters and they are called ``radio relics". Although CIZA1359 is known as the early-stage merging cluster, Source U extends about $5' \times 6'$ ($=450~{\rm kpc} \times 540~{\rm kpc}$), which is comparable in size to radio relics \citep{2012A&ARv..20...54F}.

In the above shock scenario, one may also expect radio emission from the south shock, although our observation did not find any candidate. Interestingly, \citet{2022arXiv221002145O} estimated the Mach number of the shocks and found that the north shock has a higher Mach number $\mathcal{M} = 1.7$, while the other part have a lower value of $\mathcal{M} = 1.4$. Therefore, Source U is consistent with the theoretical expectation that a shock wave with a higher Mach number forms a brighter radio emission so that it accelerates cosmic-rays more efficiently. But DSA does not work well at such low Mach number. We discuss the need for re-acceleration in section 4.2.

\subsubsection{Radio Power}

The radio power of diffuse radio emission in galaxy clusters has been studied in the literature and thus the radio power is also useful to examine whether Source U is a real emission or not. We calculate the monochromatic radio power using the following equation,
\begin{equation}
P_\nu = 4\pi D_{\rm L}^2 \int I_\nu d\Omega, 
\label{equ 1}
\end{equation}
where $D_{\rm L}(z=0.07)\sim 9.74 \times 10^{24}$~m is the luminosity distance \citep{2006PASP..118.1711W}, $I_\nu$ is the radio intensity, and $\Omega$ is the area of diffuse emission. 

\begin{figure}[tbp]
 \begin{center}
  \includegraphics[width=0.8\linewidth]{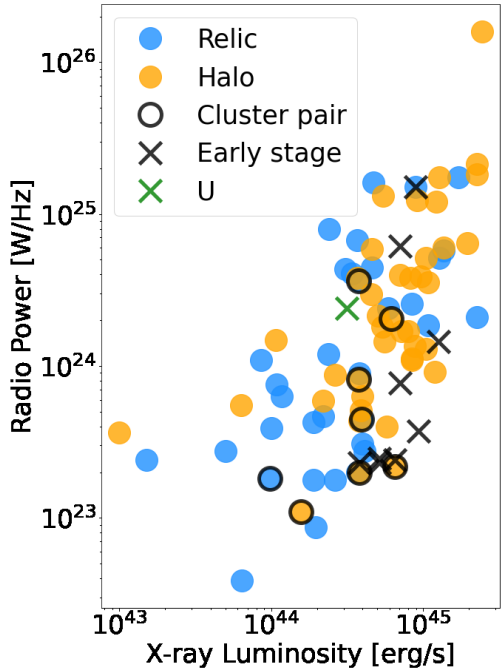} 
 \end{center}
   \caption{The relationship between radio power at 1.4~GHz and X-ray luminosity at 0.1 -- 2.4~keV of the diffuse radio sources. The green cross is data of source U obtained from this study. The blue and orange dots mean the data of the radio halo and relic, respectively, taken from \citet{2012A&ARv..20...54F}. \rev{Some cluster pairs (table~\ref{tab:6}) and early stage mergers (table~\ref{tab:7}) are shown by the black ring and black cross, respectively. } }\label{fig:radio vs xray}
\end{figure}

We integrate $I_\nu$, which is the point source subtracted intensity, within the area of the white contours in figure \ref{fig:source U} (a). To compare the monochromatic radio power in this work with those of 1.4~GHz in the literature, it was converted to the flux at 1.4~GHz from flux at 400~MHz using the spectral index of $-1.22$. The derived radio power is $P_{\rm 1.4~GHz} = 2.40 \times 10^{24}$~W Hz$^{-1}$. This is consistent with the known radio relics, halos, shown in figure \ref{fig:radio vs xray}. Therefore, there is no immediate problem in considering Source U as a diffuse radio source of a galaxy cluster.

Note that if we adopt the formula of equation (2) in \citet{2018PASJ...70...53A} to derive the upper limit of the radio power, we obtain the value broadly consistent with that derived in their work. The reason for the non-detection in ATCA can be due to the steep spectrum of Source U.

\subsubsection{Magnetic Field }

Finally, although there are uncertainties caused by theoretical assumptions, it is possible to derive magnetic-field strength, and to discuss a reality of the candidate from comparison with previous estimations. Assuming the energy equipartition between magnetic fields and cosmic-rays, the magnetic field strength can be estimated from synchrotron radiation as follows \citep{2005AN....326..414B},
\begin{equation}
B_{\rm eq} = \left\{ \frac{4 \pi (-2\alpha + 1)( K_0 + 1)I_\nu E_{\rm p}^{1+2\alpha}(\nu/2c_1)^{-\alpha}}{(-2\alpha -1)c_2({-\alpha})L\ c_4(i)} \right\} ^{1/(-\alpha+3)} ,
\label{equ:strength}
\end{equation}
where $\alpha$ denotes the spectral index of synchrotron radiation, $K_0$ is the number densities ratio of cosmic-ray nuclei to that of the electrons, $L$ is the path length of the synchrotron emitting media, $I_\nu $ is the intensities at frequency $ \nu $, and $E_{\rm p} $ is the proton rest energy. The coefficients are, $c_1 = 3e/(4\pi m_{\rm e}^3c^5) = 6.3 \times 10^{18}\ {\rm erg^{-2} s^{-1} G^{-1}}$, $c_2 = 4.56 \times 10^{-24}\ {\rm erg~G^{-1} sterad^{-1}}$, $c_4 = 1$, and $E_{\rm p} = 1.5 \times 10^{-3}\ {\rm erg}$. We adopted our best-fit value of $\alpha = - 1.22$ and $\nu = 400$~MHz, and we assumed the typical values of $L=500~{\rm kpc} \sim 1.5\times 10^{24}\ {\rm cm}$ and $K_0=100$ \citep{2017A&A...600A..18K}. We then obtained the field strength of 2.1~$\mu$G from the intensity of Source U, $0.4$~mJy~beam$^{-1}$. Such $\mu$G magnetic field is commonly found in galaxy clusters (e.g., \cite{2017A&A...600A..18K}). We note that this strength is insensitive to the parameters of equation (\ref{equ:strength}). 1 $\mu$G-order field strength is derived even when the parameters are changed. However figure \ref{fig:source U} suggests that a patchy structure exists within Source U. The structure can cause error in the estimation of the average strength in equation (\ref{equ:strength}).

We can check the field strength from an empirical radial dependence of magnetic field. Using the following equation \citep{2010A&A...513A..30B},
\begin{equation}
B(r) = B_0 \times \left( \frac{n_e(r)}{n_0} \right) ^{\eta}, n_e = n_0 \left( 1+\frac{r^2}{r_c ^2} \right) ^{-\frac{3}{2} \beta},
\label{equ:mag strength}
\end{equation}
the field strength is estimated to be $2.7~\mu$G at a relic position, $2.'5$~(225~kpc) away from the cluster centre in the north, assuming the central magnetic strength, $B_0 = 4.7~\mu {\rm G}$ , radial power-law slope, $\eta = 0.5$ \citep{2010A&A...513A..30B}. Also we adopt the central gas density, $n_0 = 2.54 \times 10^{-3}~{\rm cm^{-3}}$, the core radius, $r_c = 165~{\rm kpc}$, and the $\beta$-model parameter, $\beta = 0.67$ from \citep{2022arXiv221002145O}. Note that $B_0$ and $\eta$ are from the Coma cluster and be expected to vary with cluster. Indeed, in Abell 2382, $B_0 = 3.5~\mu {\rm G}$ and $\eta = 0.5$ are obtained \citep{2008A&A...483..699G}, and in Abell 2255, $B_0 = 2~\mu {\rm G}$ and $\eta = 0.5$ are obtained \citep{2006A&A...460..425G}. Since these value of CIZA1359 is unknown, we adopted those of the Coma cluster as representative values.

These magnetic field strengths of CIZA1359 estimated by these two independent methods are in good agreement, although they are large \rev{theoretical assumptions}. This means that the magnetic field strength of CIZA1359 is consistent with other galaxy clusters.

\subsection{Origin of Source U}

Our assessment of Source U based on its location, structure, spectrum, power, and magnetic field do not suggest that Source U is a noise. In this subsection, we discuss the origin of Source U, supposing that Source U is real.

\rev{\subsubsection{Comparison with other early-stage merging clusters}}

\rev{To understand the origin of Source U, it is useful to compare it with diffuse radio sources in early-stage merging clusters. This is because the cluster's physical properties that would be related to the origin are very different between early- and late-stage merging clusters. Here, since it is rather difficult to identify early-stage merging clusters, and there is no catalog of early-stage merging clusters yet, we look for them using the following two methods. The first is to find radio-associated cluster pairs, and the second is to make a list of well-known early-stage merging clusters.}

\rev{Radio-associated cluster pairs were searched by catalog matching. We (1) checked the coordinates of the radio-associated clusters listed in tables 1 and 3 of \citet{2012A&ARv..20...54F} using SIMBAD, and (2) cataloged if they are within the Plank's beam FWHM (7.18 arcmin) from the coordinates listed in table 1 of \citet{2013A&A...550A.134P} using TOPCAT \citep{2005ASPC..347...29T}. As a result, we found 7 radio-associated cluster pairs (table \ref{tab:6}). It should be noted that this catalog may contain not only early-stage merging clusters but also late-stage merging clusters which have close separations of subclusters. Even more, some may be random pairs which are only close in projection on the plane of the sky and have significantly different redshift, although such a case is rare because galaxy clusters sparsely exist in the Universe.}

\begin{table*}[tbp]
\tbl{Radio parameters of the cluster pairs. Columns are: (1) Name of radio associated cluster. (2) Right-ascension of the cluster. (3) Declination of the cluster. (4) Structure with radio information in column 5 and 6. (5) Logarithm of radio power at 1.4~GHz from \citet{2012A&ARv..20...54F}. (6) X-ray luminosity in the 0.1--2.4~keV band in $10^{44}$ units from \citet{2012A&ARv..20...54F}. (7) Names of cluster that pair with the one in column 1. (8) Expected merging phase that the "early" and "complex" mean the early stage which have not completed a core-crossing and the multiple merger, respectively.
}{%
\begin{tabular}{cccccccc} \hline \hline
Name & R.A. & Dec & Structure & Log~$P$(1.4) & $L_X (10^{44})$ & pair & merger phase \\
 & deg & deg &  & W/Hz & erg/sec &  &  \\
 (1) & (2) & (3) & (4) & (5) & (6) & (7) & (8) \\ \hline
A209 & 22.990  & -13.576  & halo & 24.31  & 6.17  & A222 & - \\
A399 & 44.485  & 13.016  & halo & 23.30  & 3.80  & A401 & early$^{(a)}$ \\
A401 & 44.737  & 13.582  & halo & 23.34  & 6.52  & A399 & early$^{(a)}$ \\
A2061 & 230.336  & 30.671  & relic & 23.65  & 3.95  & A2067 & - \\
A2063 & 230.758  & 8.639  & relic & 23.26  & 0.98  & MKW3s & - \\
A2256 & 255.931  & 78.718  & halo & 23.91  & 3.75  & A2271 & complex$^{(b)}$ \\
A2256 & 255.931  & 78.718  & relic & 24.56  & 3.75  & A2271 & complex$^{(b)}$ \\
A3562 & 203.383  & -31.673  & halo & 23.04  & 1.57  & A3558 & complex$^{(c)}$ \\ \hline
\end{tabular}}\label{tab:6}
\begin{tabnote}
$^{(a)}$\citet{2019Sci...364..981G}; $^{(b)}$\citet{2020MNRAS.495.5014B}; $^{(c)}$\citet{2018MNRAS.481.1055H};
\end{tabnote}
\end{table*}

\rev{Some clusters including new discoveries are well-known as early-stage merging clusters (e.g., 1E 2216.0-0401 and 1E 2215.7-0404; \cite{2019NatAs...3..838G}). Table \ref{tab:7} summarizes the information of the clusters that are believed to be early-stage merging clusters. 1E2216.0-0401 indicates a temperature jump of the ICM between the cluster pair \citep{2019NatAs...3..838G}. They suggested that the system is an early-stage merging cluster. They also reported diffuse radio sources between the cluster pairs. They concluded that they are bright AGNs affected in part by the merger shock. Abell~141 \citep{2021PASA...38...31D} also has a temperature jump and was reported to be an early-stage merging cluster. Radio structures were also found between the subclusters, but it is not possible to isolate whether they are radio bridge, relic, or halo due to lack of spatial resolution. Abell~1775 has a similar X-ray morphology like early-stage merging clusters \citep{2021A&A...649A..37B}. Sloshing or slingshot effects have been reported. Diffuse radio sources were detected and reported that their structures seem to be slingshot radio halos associated with the X-ray structure. Abell~115 clearly shows two subclusters in its X-ray morphology \citep{2020ApJ...894...60L}. No radio structure was detected between them, while a radio relic is present and implies rather a late-type merging cluster.  A3391-A3395 \citep{2021A&A...647A...3B} and A98 have radio structure that seem to associate with the head-tail galaxy \citep{2014ApJ...791..104P}.
}

\begin{table*}[tbp]
\tbl{List of well-known early-stage merging clusters. Columns (1) to (6) are the same as in table \ref{tab:6}. Column (7) is cluster pair number.
}{%
\begin{tabular}{ccccccl} \hline \hline
Name & R.A. & Dec & Radio source & Log~$P$(1.4)& $L_X (10^{44})$ & Pair \\
 & deg & deg & & W/Hz & erg/sec & \\
 (1) & (2) & (3) & (4) & (5) & (6) & (7)\\ \hline
A98 & 11.608  & 20.490  & galaxy & - & - & 1 \\ \hdashline
A115 & 13.998  & 26.321  & relic & 25.18  & 8.9 & 2 \\ \hdashline
A141 & 16.376  & -24.655  & unknown & 24.16 & 12.6$^{(d)}$ & 3 \\ \hdashline
A399 & 44.485  & 13.016  & halo & 23.36  & 3.8 & \CenterRow{4}{4} \\
A399/401 d & 44.737  & 13.582  & relic & 23.39 & 5.2$^{(e)}$  &  \\
A399/401 f & 44.737  & 13.582  & relic & 23.36  & 5.2$^{(e)}$ &  \\
A401 & 44.737  & 13.582  & halo & 23.38 & 6.5 &  \\ \hdashline
A1758 & 203.134  & 50.510  & relic & 23.57 & 9.4$^{(f)}$ & \CenterRow{3}{5} \\
A1758N & 203.134  & 50.510  & halo & 24.79 & 7.1 & \\
A1758S & 203.134  & 50.510  & halo & 23.89 & 7.1 & \\ \hdashline
A1775 & 205.482  & 26.365  & halo & - & 1.5$^{(f)}$ & 6 \\ \hdashline
A3391 & 96.564  & -53.681  & galaxy & - & 1.3$^{(g)}$ & \CenterRow{2}{7}\\
A3395 & 96.880  & -54.399  & galaxy & - & 1.3$^{(g)}$ & \\ \hdashline
1E 2215.7-0404 & 334.585  & -3.828  & galaxy & -  & 0.8$^{(h)}$ & \CenterRow{2}{8}\\
1E 2216.0-0401 & 334.673  & -3.766  & galaxy & -  & 0.8$^{(h)}$ & \\ \hdashline
CIZA1359 & 209.667  & -47.767  & relic & 24.38 & 3.1$^{(i)}$ & 9 \\ \hline
\end{tabular}}\label{tab:7}
\begin{tabnote}
$^{(d)}$\citet{1996MNRAS.281..799E}; $^{(e)}$Average value of A399 and A401; $^{(f)}$\citet{1998MNRAS.301..881E}; $^{(g)}$\citet{2009ApJ...692.1033V}; $^{(h)}$\citet{1990ApJS...72..567G}; $^{(i)}$\citet{2015PASJ...67...71K};
\end{tabnote}
\end{table*}

\rev{
Figure \ref{fig:radio vs xray} plots the radio-associated cluster pairs (table \ref{tab:6}) and the early-stage merging clusters (table \ref{tab:7}) as black circles and black crosses, respectively. We see that the distributions of them are very scattered in the relation between radio power and X-ray luminosity. CIZA1359, the green cross, appears to be inside the dispersed distribution, supporting that there is no contradiction in considering CIZA1359 to be a member of these families.
}

\rev{\subsubsection{Is Source U a radio relic?}}

\rev{We next discuss which Source U can be classified into. As broadly described in Introduction, there are some known classifications of diffuse radio emission of galaxy clusters (see e.g.,\cite{2012A&ARv..20...54F, 2019SSRv..215...16V} for reviews). We look into acceleration mechanism of cosmic-ray electrons which produce the observed radio emission of Source U. Discussion about the acceleration mechanism is helpful for the source classification.}

As discussed above, Source U seems to be associated with the northern shock with the Mach number of 1.7. Based on the standard DSA and test-particle regime, the energy spectrum of the relativistic electrons, $p$, of $n(E)dE\propto E^{-p}dE$, depends on the shock compression, $C$, as  $p = (C+2)/(C-1)$. We obtain $C \sim 1.96$ ($=(\frac{3}{4\mathcal{M}^2}+0.25)^{-1}$) for the Mach number 1.7 and give $p \approx 4.1$. If magnetic field is roughly constant over the radio source, such a power-law electron distribution will lead synchrotron emission with a spectral index $\alpha_{\mathcal{M}} = - (p-1) / 2 \approx -1.56$ for $F_{\nu} \propto \nu^{\alpha_{\mathcal{M}}}$. The observed spectral index and its error $\alpha \pm \sigma_{\rm ind} = - 1.22 \pm 0.33$ is consistent within $\sim 1$ $\sigma_{\rm ind}$ error with the estimated spectral index. This could be the case that the DSA is working for the particle acceleration at cluster shocks. 

It is, on the other hand, already known that the acceleration efficiency at a  weak shock is way too low to reproduce the observed radio luminosity (e.g., \cite{2012ApJ...756...97K,2013MNRAS.435.1061P, 2014ApJ...780..125V}). To explain this issue, several possibilities are proposed which include 1.) re-acceleration of pre-accelerated electrons (e.g., \cite{2005ApJ...627..733M, 2012ApJ...756...97K, 2016ApJ...818..204V}), 2.) shock drift accelerations (e.g., \cite{2011ApJ...742...47M, 2014ApJ...797...47G, 2014ApJ...794..153G}), 3.) other mechanisms, for instance turbulence accelerations (e.g., \cite{2015ApJ...815..116F, 2016PASJ...68...34F, 2017ApJ...840...42K}) and of course all observational systematic on both X-ray and radio side (e.g., \cite{2017A&A...600A.100A, 2017MNRAS.465.2916S, 2018MNRAS.478.2218H}). 

Indeed, there are seven compact sources in the region of Source U, with spectral indices of $-1.5$ to $-0.5$ which are comparable to the typical value of radio jets \citep{2019A&A...622A..17S}. Moreover, Source U is well aligned with the temperature jump (figure \ref{fig:source U}). Therefore, a possible origin of Source U would be the result of merger shock re-acceleration of pre-seeded cosmic-ray electrons by AGN. Note that the spectral index map (figure \ref{fig:source U} (b)) shows patchy distribution and less clear global gradient due to the aging with respect to the shock propagation direction, implying multiple seeding from these AGN.

Moreover, our no detection of diffuse radio emission associated with the southern shock with the Mach number of $1.4$ may indicate that there is a threshold of efficient (re-)acceleration at the shock age of $\sim 50$~Myr \citep{2016HEAD...1511103K}. It is thus important to examine whether theoretical models of particle acceleration can explain this marginal detection and non detection simultaneously. There are also compact radio sources in the other shock regions, and for example, the spectral indices range from $-1.5$ to $-0.5$ for six compact sources at the eastern part of the south shock. Thus, there is also a possibility of seeding cosmic-ray electrons from AGN. It is necessary to clarify whether they are member galaxies associated with CIZA1359. Those are future works of this paper.

Finally, we discuss that which of the traditional classifications for cluster diffuse emission this fits into. Source U indicates that (I) it is found in between two subclusters of an early-stage merging cluster, (II) it has a structure along the shock, (III) it possesses a relatively flat spectral index of $-1.22$, 
\rev{
and (IV) multiple non-bright radio point sources are located within the diffuse radio emission and there are no bright AGN.
}
These facts implies that Source U is a radio relic. Feature (I) is also seen in the radio bridges of early-stage merging galaxy clusters, such as Abell~399 and 401 \citep{2019Sci...364..981G} and Abell~1758 \citep{2020MNRAS.499L..11B}. However, these clusters also have radio halos and relics \citep{2018MNRAS.478..885B}, while CIZA1359 has no any other diffuse sources. Radio relics tend to be brighter than radio bridges, so that we expect which applies to CIZA1359 as well. As no other diffuse emission has been detected in CIZA1359, it is more reasonable to consider Source U as a relic than a bridge.

\citet{2019Sci...364..981G} found the formation scenario of the radio bridges between Abell 399 and 401. That is, contact of two clusters generates a shock and turbulence is excited at the post-shock region. Seed cosmic-ray electrons are re-accelerated through turbulence by the Fermi 2nd order acceleration mechanism. Such turbulence-(re)acceleration could be realized in CIZA1359 as well, though the short age of $\sim 50$ Myr prefers the direct shock acceleration by the Fermi 1st order acceleration mechanism. In other words, the radio relic candidate of CIZA1359 is a precursor of radio bridge. Even in the case, the low Mach number such as $\mathcal{M} = 1.7$ would require seeding of cosmic-rays to achieve efficient acceleration and radio emission (e.g., \cite{2019NatAs...3..838G}).

\section{Summary}

We reported the results on a SPAM-based analysis of uGMRT observations at 300--500~MHz for the early-stage merging galaxy cluster, CIZA J1358.9-4750 (CIZA1359). We found many radio sources such as a head-tail galaxies, FRII types radio lobes, AGNs, and so-called radio phoenixes or fossils. We found a diffuse radio source candidate named Source U with the flux density of $24.04 \pm 2.48$~mJy roughly along a part of the shock front found in the previous X-ray observations.

We discussed whether Source U is real or noise from several aspects of its properties. First, the location of Source U is consistent with that of the shock front. Such an association is often seen in radio relics. Second, the size is comparable to known radio relics. Interestingly, the structure of Source U coincides with the shock structure where the Mach number of the shock wave reaches its maximum value of $\mathcal{M} \sim 1.7$. Third, the relation between the radio power and the X-ray luminosity is in good agreement with that of other radio relics. And finally, the energy-equipartition magnetic-field strength, 2.1~$\mu$G, is a typical value seen in galaxy clusters and relics. The above facts favor that Source U is a real radio relic.

If Source U is a real diffuse radio source, this study confirmed that even a very weak ($\mathcal{M} \sim 1.7$) shock can accelerate cosmic-rays and emit observable radio emission. Moreover, we did not find any radio candidate at the shock with $\mathcal{M} \sim 1.4$, suggesting the existence of an acceleration-efficiency threshold around the Mach number. We suspect that seed cosmic rays were supplied by some of compact radio sources (AGN) associated with Source U and the re-acceleration was taken place at the shock.

It is important to identify the redshifts of the radio sources in order to elucidate the origin of the relic candidates. The identification is also necessary to examine whether head-tail galaxies seen in this observation interact with CIZA1359 or not. In addition, the relic candidate has a relatively steep spectrum. Therefore, future observations at lower frequencies such as 144~MHz would be promising to detect the candidate. Finally, ultimate sensitive observations with the Square Kilometre Array (SKA) will enable us to understand the details of radio sources as well as its polarization.


\begin{ack}
This work was supported in part by JSPS KAKENHI Grant Numbers, 17H01110(TA), and 21H01135(TA). AIPS is produced and maintained by the National Radio Astronomy Observatory, a facility of the National Science Foundation operated under cooperative agreement by Associated Universities, Inc. Data analysis was carried out on the Multi-wavelength Data Analysis System operated by the Astronomy Data Center (ADC), National Astronomical Observatory of Japan. RJvW acknowledges support from the ERC Starting Grant ClusterWeb 804208.
\end{ack}

\appendix 
\section*{A. Flux accuracy}

\begin{figure}[tbp]
 \begin{center}
  \includegraphics[width=\linewidth]{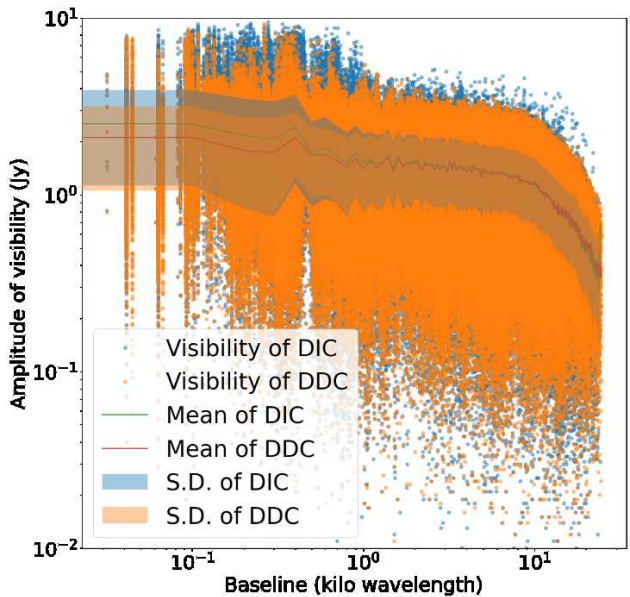} 
 \end{center}
\caption{
\rev{
Blue and orange points mean the data from DIC and DDC, respectively. The solid lines show the data averaged over 0.1 k$\lambda$ bins, and the shadows show their standard deviations.
}
}\label{fig:A03}
\end{figure}

\rev{
We confirmed the amplitude of visibility with respect to the DDC and DIC results as described in section 3.1. Figure \ref{fig:A03} shows the amplitude of visibility at each baseline length.
}

\rev{
The DDC calibration employs data over 1k$\lambda$. The amplitudes of DDC and DIC are different in each region of the figure \ref{fig:A03}. At scales below the most diffuse component (Source U), DDC has an amplitude that is about 6 \% lower than DIC.
}

\begin{figure*}[tbp]
 \begin{center}
  \includegraphics[width=0.55\linewidth]{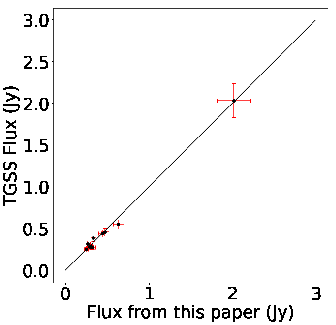} 
  \includegraphics[width=0.41\linewidth]{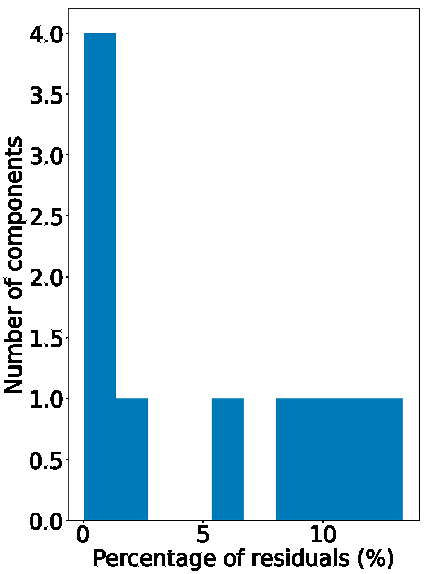} 
 \end{center}
\caption{
\rev{
{\it Left (a):} The red bar shows a 10\% error bar relative to Flux. The black line is the expected line (y=x). {\it Right (b):} The right panel shows the distance between the black line and the data divided by the flux from this paper, i.e., the percentage of error relative to flux. 
}
}\label{fig:A04}
\end{figure*}

\rev{
We further compared our results to the flux of the TIFR GMRT Sky Survey (TGSS;\cite{2017A&A...598A..78I}) to test the validity of the 10\% error in absolute flux. First, we convolved TGSS and our results to the same resolution (65~asec). Second, we performed Gaussian fitting on each component with SN$>20$ using pybdsf. Finally, we matched for each component by position using TOPCAT \citep{2005ASPC..347...29T}. The separation used for the matching was the same value as the resolution, and since TGSS is a 150~MHz band, it was re-scaled using $\alpha = -0.8$. The resulting flux relationship is shown in the figure \ref{fig:A04}. These results mean that the error in flux of this paper is usually well within 10\%. The minimum flux in the figure \ref{fig:A04} depends on the noise level of the TGSS.
}

\section*{B. Spectral index}

\begin{figure}[tbp]
 \begin{center}
  \includegraphics[width=\linewidth]{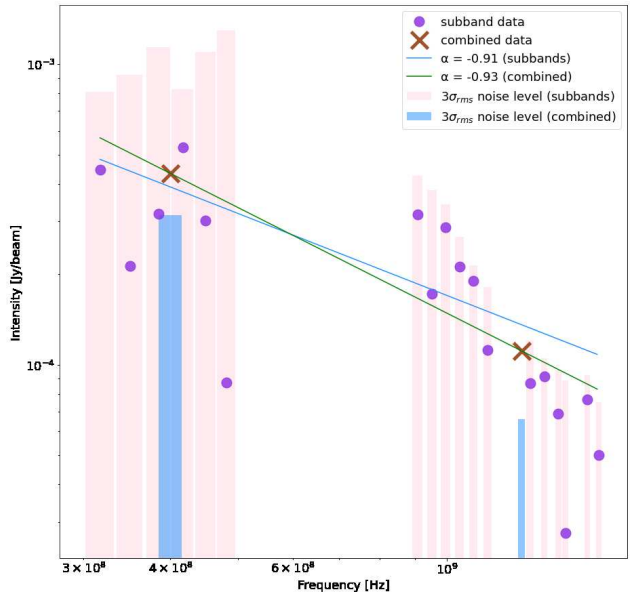} 
 \end{center}
\caption{
\rev{
Example of radio spectrum in Source U. The points marker shows the data for each frequency channel. The cross is combined data from uGMRT and MeerKAT, respectively. The solid lines show the best-fit power-law models, where the blue line considers all data for each frequency channel. The green line is the result of fitting using combined data from uGMRT and MeerKAT, respectively. The pink and blue bars indicate the $3\sigma_{rms}$ of each channel and the combined data, respectively.
}
}\label{fig:spec}
\end{figure}

 An example of a spectral fit is shown in figure \ref{fig:spec}. We examine in this section whether the spectral index of Source U is due to noise behavior.
 
\begin{figure*}[tbp]
 \begin{center}
  \includegraphics[width=0.9\linewidth]{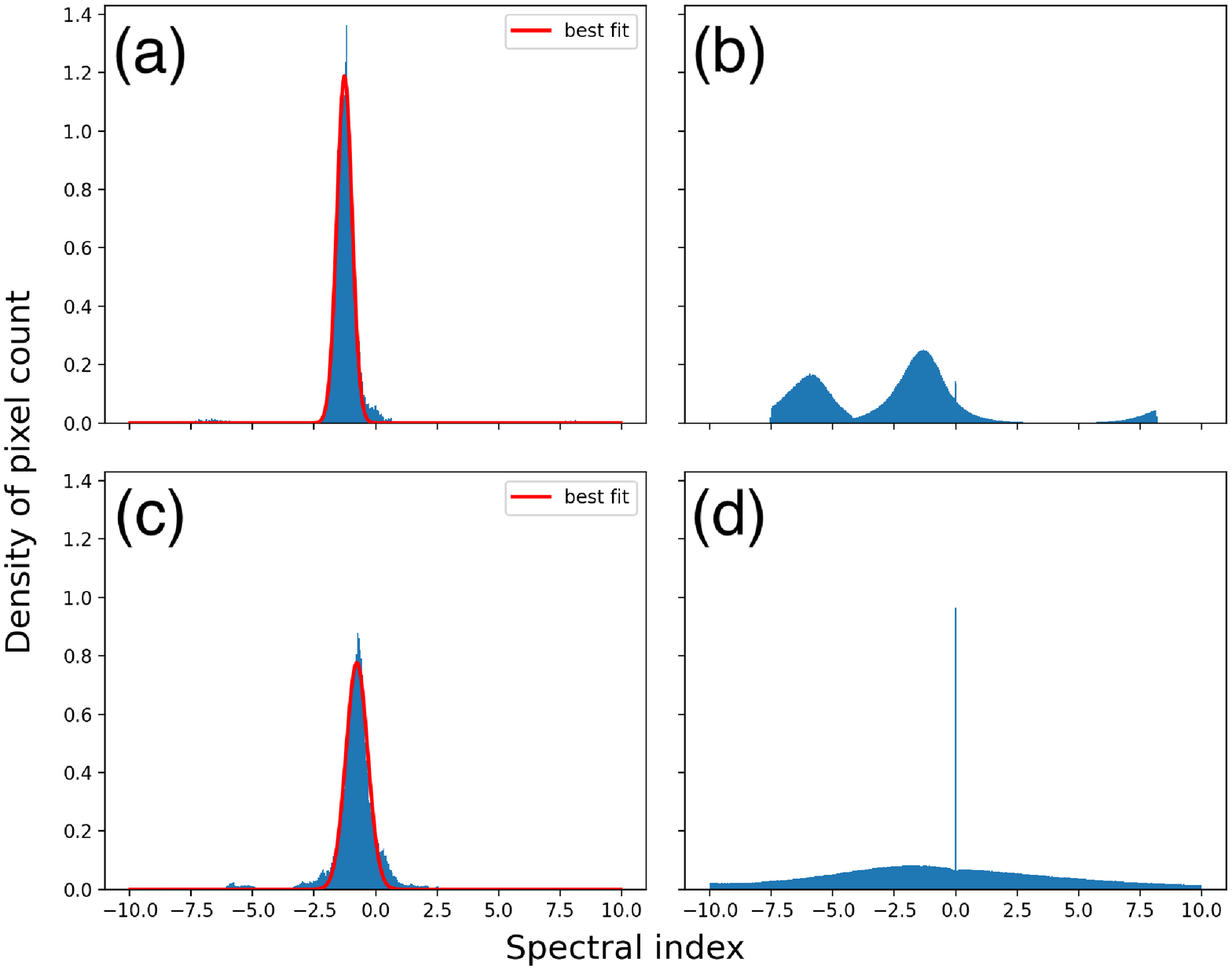} 
 \end{center}
\caption{Histograms of spectral indices. This is normalized so that the total area is unity. The histograms were sampled between -10 and 10 with even intervals of 0.05 in spectral index axis. {\it Top left (a)}: Spectral index histogram of Source U. The blue bars show the data. The red line show the best Gaussian fit. {\it Top right (b)}: Spectral index histogram of all the pixels within the image. The data of uGMRT and MeerKAT are both taken for the power-law fit. {\it Bottom left (c)}: Same as (b) but for the pixels with an intensity greater than $0.4$~mJy~beam$^{-1}$ at the smoothed uGMRT image which mean the white contours in figure \ref{fig:radio image2}. {\it Bottom right (d)}: Same as (b) but the spectral index was calculated only with uGMRT data.}\label{fig:hist}
\end{figure*}

Figure \ref{fig:hist}(a) show the histogram of the spectral indices for the pixels within the Source U. We obtained the average spectral index of $-1.23 \pm 0.27$, the error means the standard deviation of the histogram. However, we need a careful assessment of the spectral index for marginal sources like Source U, because of the significant contribution of noise for the spectrum index fitting.

\begin{figure*}[tbp]
 \begin{center}
  \includegraphics[width=0.49\linewidth]{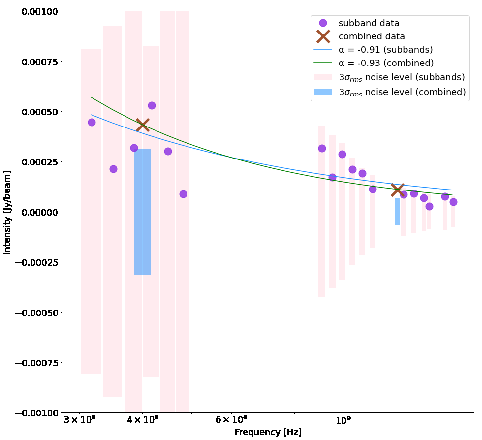} 
  \includegraphics[width=0.49\linewidth]{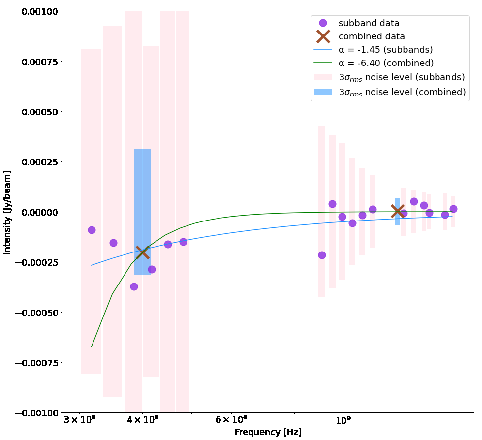} 
 \end{center}
\caption{ {\it Left (a):} Same as figure \ref{fig:spec}, but the vertical axis is a linear. {\it Right (b):} Same as the left panel, but it is the spectrum of pixel where noise is dominant.}\label{fig:ape2}
\end{figure*}

We have checked the spectral indices of all pixels in the image and found that they distributes with three peaks at 0, $-0.7$ and $-6.8$ as shown in figure \ref{fig:hist}(b). The 0-centered peak is thought to be derived from spectral fitting by bad pixels, which the initial value of 0 remains as it is because the fitting does not converge; indeed the 0-centered peak disappears in the histograms of the pixels brighter than $0.5$~mJy~beam$^{-1}$, i.e. the pixels possessing high signal-to-noise ratio (SN $\gtrsim 5$), which has peak at $-0.71 \pm 0.47$ (figure \ref{fig:hist}(c)). The other, $-0.7$- and $-6.8$-centered peaks are thought to be derived from spectral fitting between noise floors in the data of uGMRT and MeerKAT\footnote{The image rms noise of uGMRT is larger than that of MeerKAT. When given error values of uGMRT and MeerKAT with the same sign, they make an artificial negative spectral index up. In our data, this artificial index is $-0.7$}. The $-0.7$-centered peak is consistent with the values estimated from the noise distribution such as shown in figure \ref{fig:spec}. The $-6.8$-centered peak occurs frequently when fitting negative and positive noise (see figure \ref{fig:ape2} (b)). Indeed, the index histogram calculated from only uGMRT data (figure \ref{fig:hist}(d)) does not show the $-0.7$- and $-6.8$-centered peaks which has no clear peak other than 0-centered. 

We also show the spectra in linear space in figure \ref{fig:ape2}. The left panel (a) of the figure \ref{fig:ape2} shows the spectrum of the same pixel as in figure \ref{fig:spec}, and the right panel (b) of the figure \ref{fig:ape2} shows the spectrum of the noise pixel at the outer edge of the field of view. The figure \ref{fig:ape2} (a) tends to have predominantly positive data compared to noise pixel (figure \ref{fig:ape2} (b)), which suggests that Source U is different from the noise trend. Also, the green line in figure \ref{fig:ape2} (b) shows that the spectral index is very steep because of the fitting with negative 400~MHz data and positive 1280~MHz data.

The above results support that it is more favor to consider that Source U as a real radio source. The value, $- 1.23 \pm 0.27$, is consistent with the spectral index of Source U inferred from the flux density.

\end{document}